\documentclass[sigconf,balance=false]{acmart}
\usepackage{popets}

\usepackage{popets}
\usepackage{booktabs}
\usepackage[font={small}]{caption}
\usepackage{float}
\usepackage[shortlabels]{enumitem}
\usepackage{listings}
\usepackage{subcaption}

\microtypecontext{spacing=nonfrench}

\usepackage{enumitem}

\newlist{questions}{enumerate}{2}
\setlist[questions,1]{label=\textbf{RQ\arabic*.},ref=RQ\arabic*, left=0em, align=left}
\setlist[questions,2]{label=(\alph*\tex),ref=\thequestionsi(\alph*)}

\colorlet{shadecolor}{gray!35}

\setcopyright{popets}
\copyrightyear{YYYY}

\acmYear{YYYY}
\acmVolume{YYYY}
\acmNumber{X}
\acmDOI{XXXXXXX.XXXXXXX}
\acmISBN{}
\acmConference{Proceedings on Privacy Enhancing Technologies}
\begin{document}

\title[Exercising the CCPA Opt-out Right on Android]{Exercising the CCPA Opt-out Right on Android:\\ Legally Mandated but Practically Challenging}

\author{Sebastian Zimmeck}
\affiliation{%
  \institution{Wesleyan University}
  \city{Middletown}
  \state{Connecticut}
  \country{USA}}
\email{szimmeck@wesleyan.edu}
  
\author{Nishant Aggarwal}
\affiliation{%
  \institution{Wesleyan University}
  \city{Middletown}
  \state{Connecticut}
  \country{USA}}
\email{naggarwal@wesleyan.edu}
  
\author{Zachary Liu}
\affiliation{%
  \institution{Wesleyan University}
  \city{Middletown}
  \state{Connecticut}
  \country{USA}}
\email{zliu01@wesleyan.edu}

\author{Sage Altman}
\affiliation{%
  \institution{Wesleyan University}
  \city{Middletown}
  \state{Connecticut}
  \country{USA}}
\email{jaltman01@wesleyan.edu}

\author{Konrad Kollnig}
\affiliation{%
  \institution{Maastricht University}
  \city{Maastricht}
  \state{}
  \country{Netherlands}}
\email{konrad.kollnig@maastrichtuniversity.nl}

\renewcommand{\shortauthors}{Zimmeck et al.}

\begin{abstract}
Many mobile apps' business model is based on sharing user data with ad networks to deliver personalized ads.
The California Consumer Privacy Act (CCPA) gives California residents a right to opt out.
In two experiments we evaluate to which extent popular Android apps enable California residents to exercise their right.
In our first experiment---manually exercising the right via app-level UIs---we find that only 48 out of 100 apps implement a respective setting, which suggests that CCPA opt-out right compliance on the Android platform is generally low.
In our second experiment---automatically exercising the opt-out right by sending Global Privacy Control (GPC) signals---we find for an app dataset of $1,811$ apps that GPC is largely ineffective.
While we estimate with 95\% confidence that 62\%--81\% of apps in our app dataset must respect the CCPA opt-out right, many apps do not do so.
Our evaluation of disabling apps' access to the AdID---which is technically not intended for exercising the CCPA opt-out right but could be practically effective---does not change our conclusion.
For example, when sending GPC signals and disabling apps' access to the AdID, 338 apps still had the \texttt{ccpa status} of the ad network Vungle set to \texttt{opted\_in} while only 26 had set it to \texttt{opted\_out}.
Overall, our results suggest a compliance gap as California residents have no effective way of exercising their CCPA opt-out right on the Android platform; neither at the app- nor at the platform-level.
We think that re-purposing the Android AdID setting as an opt-out right setting with legal meaning under the CCPA and other laws could close this gap and improve users' privacy on the platform significantly.
\end{abstract}


\maketitle

\section{Introduction}
\label{introduction}

\begin{figure}
    \centering
    \includegraphics[width=0.26\textwidth]{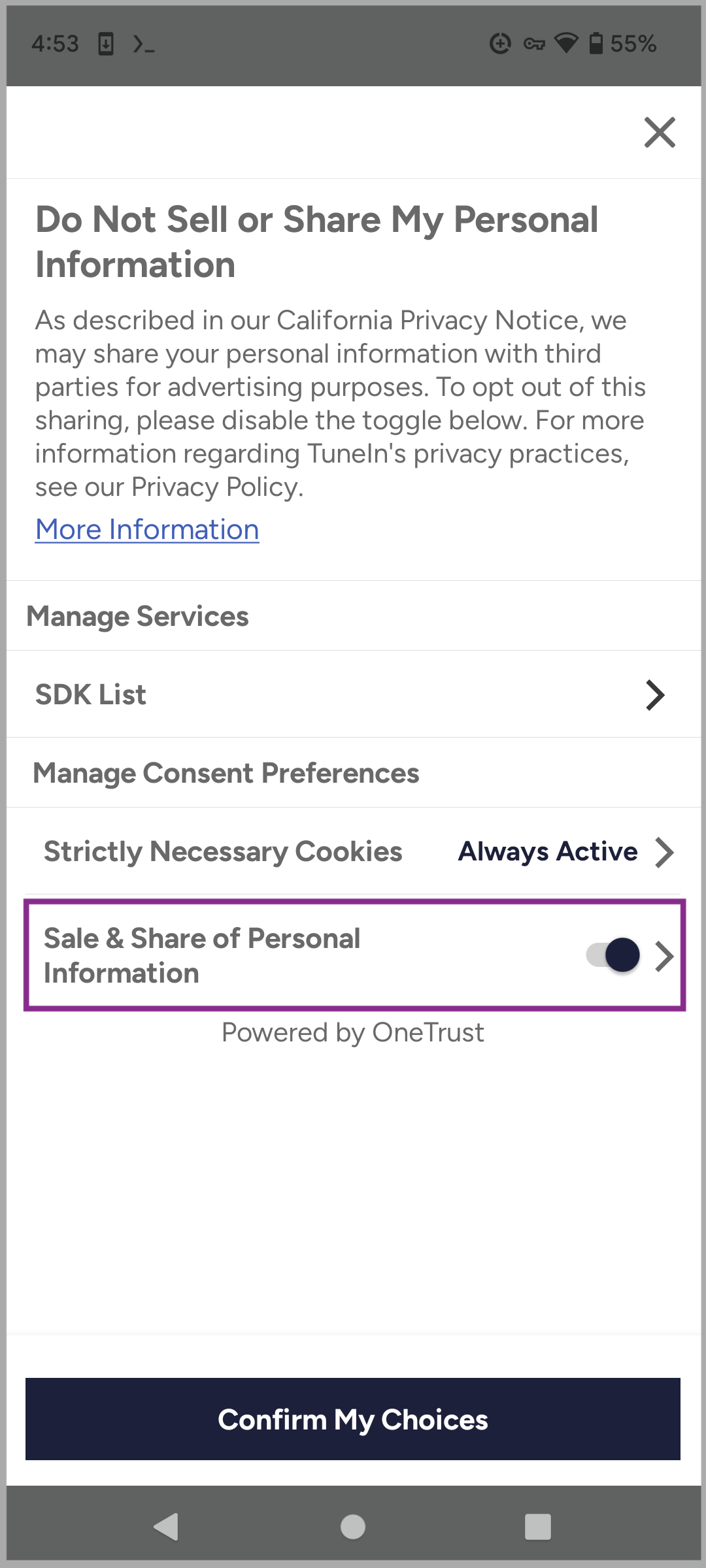}
    \caption{The \textit{TuneIn Radio: Music \& Sports} app~\cite{TuneIn} allows California residents to opt out of the selling and sharing of personal information per the CCPA.}
    \Description{The TuneIn Radio: Music \& Sports Radio app for Android allows California residents to opt out of the selling and sharing of personal information per the CCPA}
    \label{fig:opt-out-right-screenshot}
\end{figure}

The business model of many mobile apps is based on advertising.
Many apps rely on personalized advertising and share personal information with third-party ad networks, data brokers, and other companies in the online ad industry.
However, such \textit{cross-context behavioral advertising} is often not transparent and exposes users to privacy risks without control over the sharing of their data~\cite{Privacy-in-targeted-advertising-on-mobile-devices}.
The California Consumer Privacy Act (CCPA) gives California residents a right to opt out of the selling and sharing of their personal information, including cross-context behavioral advertising.
Mobile apps that are subject to the CCPA opt-out right are required to provide California residents functionality to exercise their right.
Figure~\ref{fig:opt-out-right-screenshot} shows an opt-out screen for one such app.
In this study we evaluate to which extent popular apps on the Android platform enable California residents to exercise their CCPA opt-out right, how they can do so, and which effect, if any, exercising the right has.
We focus on the Android platform as its apps have been shown to share personal information more broadly than their counterparts on the iOS platform~\cite{Kollnig_2022}.

While app-level opt-out UIs, as shown in Figure~\ref{fig:opt-out-right-screenshot}, are one way to exercise the opt-out right, they are not the only way.
The Office of the California Attorney General is also enforcing the opt-out right as expressed via Global Privacy Control (GPC)~\cite{CalAGCCPAEnforcementSephora,CalAGCCPAEnforcementHealthline}, a technical specification for opting out~\cite{GPC}.
Per the CCPA, a ``business shall treat [...] opt-out preference signal[s] as a valid request to opt-out of [the] sale/sharing [...] for [a] browser or device.~\cite{ccparegs}''\footnote{CCPA Regulations \S7025(c)(1).}
Compliance with GPC is required in California since January 2021~\cite{XavierBecerraGPCTweet2}.
However, achieving compliance is challenging for app operators since Android currently does not have a platform-level GPC setting.\footnote{The same is true for iOS, other mobile platforms, IoT platforms, and, broadly, any other platform beyond the web.}
Thus, the effectiveness of GPC on Android is unclear and has not been studied in previous research.
We see our study as a GPC requirement evaluation of the Android platform: without platform-level GPC support, to which extent do Android apps respect California residents' opt-out rights under the CCPA?

Despite the lack of platform-level GPC support, there is already an opt-out setting on Android.
While not intended for exercising legal opt-out rights, if a user has opted out using the AdID setting, Google prohibits app operators to use the AdID for creating user profiles for advertising purposes or for targeting users with personalized advertising~\cite{PlayConsoleHelpAds}.
Google requires that all apps uploaded or published to the Google Play store must generally use the AdID in lieu of any other device identifier for any advertising purposes~\cite{GoogleAdID}.
Thus, as part of our evaluation we also study the effect of disabling the AdID.
We explore whether opting out using the AdID setting effectively results in exercising the CCPA opt-out right.

We evaluate the effectiveness of opting out on Android per the CCPA with two experiments and address the following research questions:

\begin{questions}
    \item \textbf{UI Opt-out Evaluation}: For a set of 100 popular Android apps, do those apps enable California residents to opt out per the CCPA, how can they do so, and which effect, if any, does exercising the opt-out right have? (Section~\ref{UI-Opt-Out-Evaluation})
    \item \textbf{GPC Opt-out Evaluation}: For a set of 1,811 popular Android apps, what effect does sending GPC signals have for opting out per the CCPA? What effect does disabling the AdID have? (Section~\ref{GPC-Opt-Out-Evaluation})
\end{questions}

By addressing these questions, we aim to advance legally mandated opt-out rights on mobile platforms.
More broadly, we view our study as a contribution towards giving users more control over the sharing of their data on mobile platforms.
The Android and iOS platforms are currently in a transitional phase.
Mobile app operators are required to respect users' GPC signals as valid means of exercising opt-out rights in jurisdictions in which GPC is in effect.
However, they lack platform-level support in form of native Android and iOS GPC settings and respective APIs.
To that end, we proposes to re-purpose the current Android AdID setting as a platform-level opt-out setting with legal meaning (Section~\ref{discussion}).

Addressing our research questions requires a combination of legal and technical analysis.
In the following, we will discuss related work (Section~\ref{related-work}), provide a brief background discussion on the application of the CCPA opt-out right on mobile platforms (Section~\ref{legal-background}), and describe our methodology for detecting apps that are non-compliant with the right (Section~\ref{methodology}).
Based on this groundwork we will then address our research questions and discuss the practical implications of our findings for the Android platform.

\section{Related Work}
\label{related-work}

Our work builds on previous studies on the dynamic analysis of mobile apps' data practices and privacy compliance (Section~\ref{Dynamic-Analysis-of-Apps-Privacy-Compliance}), opting out of ad tracking in mobile apps (Section~\ref{Opting-Out-of-Ad-Tracking-in-Mobile-Apps}), and privacy preference signals (Section~\ref{Privacy-Preference-Signals}).

\subsection{Dynamic Analysis of Mobile Apps' Data Practices and Privacy Compliance}
\label{Dynamic-Analysis-of-Apps-Privacy-Compliance}

Our study makes use of dynamic analysis of mobile apps for purposes of privacy compliance analysis.
Early studies in this research area focused on modifying mobile operating systems to observe personal data traversing through the systems. 
Enck et al.'s TaintDroid~\cite{enck_taintdroid_2010} falls in this category for Android as does Agarwal's and Hall's ProtectMyPrivacy~\cite{agarwal_protectmyprivacy_2013} for iOS.
While this operating system-focused approach can give highly accurate insights into apps' data flows, the growing complexity of mobile operating systems makes it increasingly challenging.
Thus, more recent studies tend to focus on the capture and analysis of apps' network traffic~\cite{Ren:2016:RRC:2906388.2906392,song_privacyguard_2015,le_antmonitor_2015,236300}, which is also our approach.
In the closest work to ours, Kollnig et al. performed a dynamic analysis of the network traffic of 1,297 Android apps and found that while most engaged in third-party tracking, only few obtained user consent before doing so~\cite{kollnig2021_consent_analysis}.
While they selected their apps from the Google Play store randomly, we selected apps based on Google's lists of popular apps as those are most likely to rely on an ad-based business model that implies selling of personal information or sharing such with third parties (Section~\ref{The-CCPA's-Opt-out-Right}).
Nguyen et al.~\cite{274614,nguyen_freely_2022} extended the work of Kollnig et al. and found that 24,838/86,163 (29\%) apps sent personal data to data controllers without the user's prior explicit consent.
While previous studies focused on EU law, we focus on the CCPA.

\subsection{Opting Out of Ad Tracking in Mobile Apps}
\label{Opting-Out-of-Ad-Tracking-in-Mobile-Apps}

In many cases, the lack of consent is the result of a coordination failure.
Specifically, most third parties rely on app operators to obtain consent on their behalf but in many cases fail to put in place robust measures to ensure operators can actually do so~\cite{kollnig2021_consent_analysis}.
Consent Management Platforms (CMPs), such as OneTrust~\cite{OneTrustMobile}, provide software integrations to assist operators in obtaining consent.
For Apple devices, the platform itself includes a consent setting.
With iOS 14.5 Apple introduced mandatory choice prompts.
In doing so, Apple deviated from its earlier implementation of the ad tracking setting as an opt-out mechanism similar to that implemented on Android since 2013.
However, Apple is facing challenges to enforce its App Tracking Transparency framework~\cite{10.1145/3531146.3533116}.
Across platforms, there are also usability challenges.
Examining 100,000 Android apps, Chen et al. found that 36.3\% of privacy settings identified in these apps are hard to find due to unclear user interface designs, which impact users' privacy because 82.2\% of the observed settings are set to disclose data by default~\cite{DBLP:conf/sp/ChenZZXZFYST00Z19}.

Users often do not know which data is collected from them and by whom~\cite{zimmeckEtAlPrivacyPioneer2024}.
Industry concentration has significantly increased the breadth and depth of data accessible to individual companies engaging in data collection~\cite{10.1145/3176246}.
For a long time mobile platforms have had dedicated identifiers to enable third-party tracking. 
On iOS, Apple rolled out the Identifier for Advertisers, alongside an opt-out setting from ad tracking, in 2012~\cite{IDFA}. 
Google quickly followed suit for Android by rolling out a similar mechanism in 2013~\cite{IDFA}.
These initiatives had the goal of moving away from using permanent device identifiers, such as the Unique Device Identifier on iOS or the Android\_ID on Android, for advertising purposes.
Permanent identifiers cannot be reset by users and, thus, entail identification risk.
In the wake of these developments, various self-help tools for blocking ad tracking have emerged~\cite{le_antmonitor_2015,greene_platform_2018,kollnig2022_app}.

To exercise their legal opt-out rights for various US jurisdictions, users can use Consumer Reports' Permission Slip app, which has a human in the loop for performing some opt-out tasks~\cite{ConsumerReportsPermissionSlip}.
Further, OptOutCode is a platform-agnostic proposal to automate opting out based on users' prefixing the names of their devices with ``0\$S'' for ``do not (0) sell (\$) or share (S)'' personal information~\cite{OptOutCode}.\footnote{This idea is similar to Google's opt-out solution for Google Location services access points~\cite{GoogleLocationServicesOptOut}: ``To opt out, change the SSID (name) of your Wi-Fi access point (your wireless network name) so that it ends with `\_nomap.' For example, if your SSID is `12345,' you would change it to `12345\_nomap.'''}
The Digital Advertising Alliance, an ad industry association, provides the AppChoices app~\cite{YourAppChoices}.
However, despite these tools, privacy choices have historically been more limited on mobile platforms compared to the web. 
This development is the result of the lag of privacy mechanism implementations on mobile platforms.

\subsection{Privacy Preference Signals}
\label{Privacy-Preference-Signals}

Recently, privacy preference signals have re-emerged and present an increasingly viable opt-out method~\cite{DBLP:journals/corr/abs-2106-02283}.
However, various legal, technical, usability, and accountability challenges persist~\cite{0623945ec14640708544dbb3c5e62444}.
Automating privacy choices via privacy preference signals has been tried before.
The Platform for Privacy Preferences Project (P3P) was an early signal to help people understand data practices and automate opt-outs and other privacy choices based on machine-readable privacy policies~\cite{p3p,p3p11}. 
Advanced Data Protection Control (ADPC) follows in P3P's footsteps and is developed for EU law~\cite{ADPC}.
Perhaps, the most well-known signal, Do Not Track (DNT)~\cite{DNT}, was developed from 2009 as a binary signal for users to express their opt-out of tracking towards websites and other online services~\cite{CalOPPA}.
However, DNT, just as the other signals, was never broadly adopted because no law mandates compliance.
Recently, a decision of the Landgericht Berlin~\cite{LGBerlinDNT}---requiring LinkedIn to respect DNT signals---revitalized DNT.
However, so far the decision remains an outlier.
As a successor to DNT, GPC~\cite{GPC} is currently emerging as the standard for exercising opt-out rights.
Thus, in our study we focus on GPC.

GPC is a binary signal with which users can express their Do Not Sell Or Share preference~\cite{GPC}.
Most importantly, compliance with this preference is required under the CCPA and has been enforced by the Office of the California Attorney General~\cite{CalAGCCPAEnforcementSephora,CalAGCCPAEnforcementHealthline}.
Our GPC analysis is based on the methodology developed by Zimmeck et al. for the web~\cite{zimmeckEtAlGPC2023}, which we adapt for use on the Android platform. 
The basic idea is to inject GPC headers into all apps' outgoing HTTP requests to detect changes in app behavior that could indicate whether the opt-out via GPC is respected or not.
For example, we evaluate whether GPC signals are propagated to third parties by measuring respective changes in their US Privacy Strings~\cite{IABUSPrivacyString}.
If apps do not notify ad networks and other third parties of California residents' opt-out preferences by setting privacy flags, such omission indicates a CCPA violation~\cite{zimmeckEtAlGPC2023}.
In addition to measuring the effect of sending GPC signals, we also measure the effect of disabling the AdID.

\section{Legal Background}
\label{legal-background}

Over the past few years many countries and various states in the US have adopted new privacy laws. 
In our study we focus on California's CCPA and its implementation via the CCPA Regulations.

\subsection{The CCPA Opt-out Right}
\label{The-CCPA's-Opt-out-Right}

The CCPA gives consumers, i.e., California residents,\footnote{CCPA \S1798.140(i).} a right to opt out.\footnote{CCPA \S1798.120.}
The right does not apply to residents of other states or countries.
Specifically, a consumer has the right ``at any time, to direct a \textit{business} that \textit{sells or shares personal information} about the consumer to third parties not to sell or share the consumer's personal information.''\footnote{CCPA \S1798.120(a) (emphasis added).} 
Whether an app operator is running a ``business'' depends on annually buying, selling, or sharing the personal information of 100,000 or more consumers or households.\footnote{Alternatively, thresholds of annual gross revenues in excess of 25 million dollars or deriving 50\% or more of annual revenues from selling or sharing consumers' personal information can be used to determine whether an entity qualifies as a ``business'' per CCPA \S1798.140(d)(1).}
For this determination Samarin et al. used the number of app installs~\cite{samarin2023lessons}. 
We rely on app operators' self-declaration of CCPA applicability in their privacy policies (Sections~\ref{UI-Opt-out-Evaluation} and~\ref{GPC-and-AdID-Opt-Out-Evaluation}), which represents a lower bound since there may be additional apps that are subject to the CCPA but whose operators do not declare so.
This approach also has the advantage of identifying a business across apps.
For example, an operator may have multiple apps, each of which does not reach the threshold for CCPA applicability on its own.

For the opt-out right to be applicable, the app operator must also ``sell'' or ``share'' personal information.
``Selling'' means to communicate a consumer's personal information to a third party for monetary or other valuable consideration.\footnote{CCPA \S1798.140(ad)(1).} 
``Sharing'' means to communicate a consumer's personal information to a third party for cross-context behavioral advertising.\footnote{CCPA \S1798.140(ah)(1).}
The concept of ``sharing'' was introduced to the CCPA by the California Privacy Rights Act, an amendment to the CCPA, to remove any doubt that the CCPA applies to cross-context behavioral advertising~\cite{TrueVaultSellvsShare}.
Evidence for selling or sharing can consist of dedicated CCPA opt-out of sale or share links, respective opt-out implementations, or statements of applicability of the CCPA opt-out right in privacy policies~\cite{DBLP:journals/corr/abs-2009-07884}.
As the enforcement actions of the Office of the California Attorney General show~\cite{CalAGCCPAEnforcementHealthline}, the integration of third-party libraries for ads personalization falls squarely under the definitions of selling and sharing.
Thus, for our GPC opt-out evaluation we rely on an app's integration of an ad network that discloses in its privacy policy that it is ``buying'' or ``collecting'' personal information or ``selling'' or ``sharing'' such (Section~\ref{GPC-and-AdID-Opt-Out-Evaluation}).

\subsection{Methods for Opting Out}

The CCPA Regulations outline several methods through which businesses must allow consumers to exercise their opt-out right.\footnote{CCPA Regulations \S7026(a).} 
As a general guideline, they require businesses to consider the methods by which they interact with consumers, the manner in which they collect the personal information that they make available to third parties, available technology, and ease of use by the consumer when determining which methods consumers may use to submit requests to opt out of selling and sharing with at least one method reflecting the manner in which the business primarily interacts with the consumer.

For websites the CCPA requires businesses to provide a ``clear and conspicuous link on the business' internet homepages, titled `Do Not Sell or Share My Personal Information,' to an internet web page that enables a consumer, or a person authorized by the consumer, to opt out of the sale or sharing of the consumer's personal information.''\footnote{CCPA \S1798.135(a)(1).}
In this regard, cookie banners or cookie controls, by themselves, are insufficient because selling and sharing practices are not identical to cookie practices.\footnote{CCPA Regulations \S7026(a)(4).}
While there may be some overlap---for example, disabling targeting cookies may also reduce selling and sharing of personal information---the CCPA requires a dedicated method for submitting requests to opt out of the selling and sharing of personal information.

In addition to links, the CCPA Regulations mandate that businesses respect opt-out preference signals: ``A business that collects personal information from consumers online shall, at a minimum, allow consumers to submit requests to opt-out of sale/sharing through an opt-out preference signal and at least one of the following methods: an interactive form accessible via the `Do Not Sell or Share My Personal Information' link, the Alternative Opt-out Link, or the business’s privacy policy if the business processes an opt-out preference signal in a frictionless manner.''\footnote{CCPA Regulations \S7026(a)(1). The ``Alternative Opt-out Link'' means a link combining the two separate ``Do Not Sell or Share My Personal Information'' and ``Limit the Use of My Sensitive Personal Information'' links into one ``Your Privacy Choices,'' or, ``Your California Privacy Choices'' link per CCPA Regulations \S7001(b) and \S7015.}
The idea is that businesses should (1) prominently notify consumers of the selling and sharing of personal information and how they can opt out and (2) process their opt-out preferences via automated opt-out preference signals.
If a business processes opt-out preference signals in a frictionless manner, it is not required to provide any opt-out links.\footnote{CCPA Regulations \S7025(e).}
Businesses are required to treat opt-out preference signals as a valid request to opt out of the sale or sharing of personal information for a browser or device and any consumer profile associated with that browser or device, including pseudonymous profiles.\footnote{CCPA Regulations \S7025(c)(1).}

\subsection{Opting Out via GPC}

The most important privacy preference signal for purposes of the CCPA is GPC, which is a draft standard at the W3C~\cite{GPC}.
GPC signals are transmitted from users' clients to servers via request headers. 
Servers can also query a user agent's GPC status via the GPC DOM property. 
GPC can be set to \texttt{1} (header) or \texttt{true} (DOM property) to express a user's opt-out as defined by applicable law in the user's jurisdiction. 
Thus, GPC can be used in multiple jurisdiction and will have the meaning that a particular law or regulation attaches to it.
GPC aims to enable users to exercise their opt-out rights beyond individual websites and apps for sets of sites or apps.

Recent studies evaluating the usability of GPC suggest that most users understand its purpose and that they would turn it on if it were more broadly available~\cite{zimmeckEtAlGPC2023,zimmeckEtAlGPC2024}.
Its rate of adoption by website operators appears to increase slowly and steadily estimated at a few percentage points per year~\cite{hausladenEtAlGPCWeb2025}. 
Compliance with GPC is required in California (since January 2021)~\cite{XavierBecerraGPCTweet2}, Colorado (since July 2024)~\cite{GPCColorado}, Connecticut~\cite{ConnecticutDataPrivacyActFAQs} (since January 2025), and New Jersey~\cite{NewJerseyDataPrivacyLawFAQs} (since July 2025). 
Notably, in these jurisdictions the opt-out right and GPC are not limited in their applicability to websites but apply to mobile apps as well.

\subsection{Opting Out via GPC on Mobile Apps}

The Office of the California Attorney General brought various enforcement actions against mobile app operators for not respecting consumers' opt-out right~\cite{CCPAEnforcementCaseExamples,CalAGCCPAEnforcementSephora,CalAGCCPAEnforcementHealthline}.
Consumers have a right to opt out on every ``device,'' which the CCPA defines as ``any physical object that is capable of connecting to the internet, directly or indirectly, or to another device.''\footnote{CCPA \S1798.140(o).}
Thus, the International Association of Privacy Professionals, a nonprofit and non-advocacy membership association of privacy professionals, recommends for mobile apps that ``[a]t a minimum, mobile apps should [...] include a functional `Do Not Sell or Share My Personal Information' link in the app [...]''~\cite{IAPPMobileApps}. 
However, opting out via GPC on mobile devices presents a challenge. 

The core problem is the lack of a native and platform-level GPC setting on Android.\footnote{For that matter, such settings are lacking on any platform but the web.}
This technical gap is in conflict with the CCPA Regulations, which mandate that ``[t]hrough an opt-out preference
signal, a consumer can opt-out of sale and sharing of their personal information with all businesses they interact with online \textit{without having to make individualized requests with each business}.''\footnote{CCPA Regulations \S7025(a) (emphasis added).}
This discrepancy between the law and the technical reality raises crucial questions: How are app operators expected to facilitate GPC opt-out signals without platform-level support.
Do they even anticipate receiving such signals? 
On a fundamental level, as mobile apps are using the web for their client--server communication, including for ad-serving, they can generally process GPC signals in the same way as websites.
Thus, our study investigates to which extent this is actually the case.

\section{Methodology}
\label{methodology}
\label{dynamic-analysis-procedure}
\label{limitations}

We study the effectiveness of exercising the CCPA out right via Android apps' user interfaces and sending GPC headers.

\begin{figure*}[!t]
    \centering
    \includegraphics[width=\textwidth]{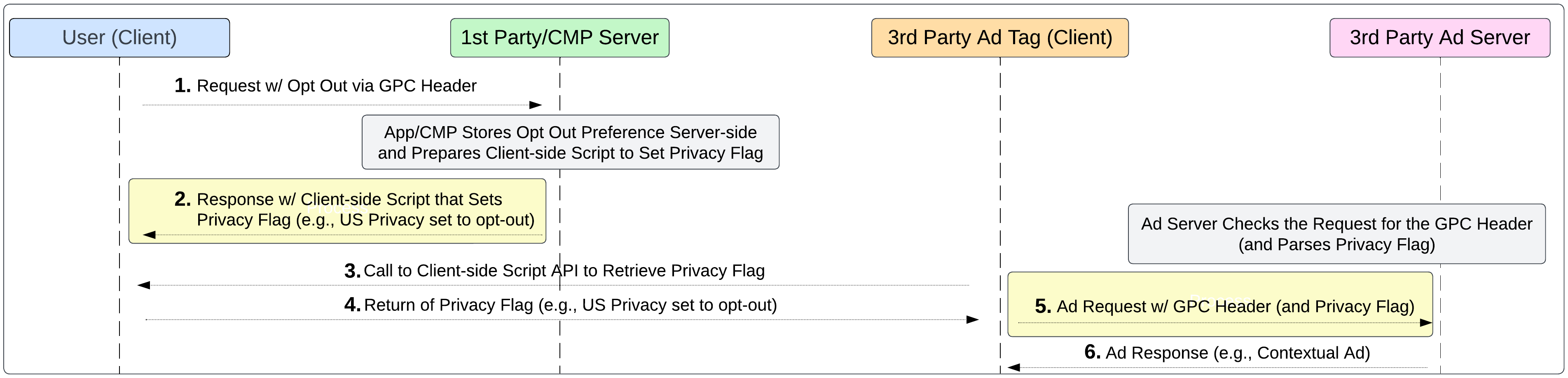}
    \caption{Exercising the opt-out right via GPC. 1. An HTTP request from the user's device is sent with a GPC header. By sending this header the user is exercising the opt-out right. 2. The HTTP response contains a script for setting a privacy flag client-side. By identifying such a privacy flag and its value we determine whether the opt-out has been respected. App operators can implement the logic for processing privacy flags themselves or use the library of a CMP. 3--4. A third-party ad network queries the privacy flag via its ad tag that the app operator integrated in its app. 5. The flag may be attached to ad requests from the user to the ad network, which is the second point that we use to determine whether the opt-out has been respected. The ad network can also act directly upon receiving GPC signals, which are attached to all requests as headers. 6. The ad network responds to ad requests from opted out users with an ad not based on sold or shared personal information.}
    \Description{Exercising the opt-out right via GPC.}
    \label{fig:GPC_Android_Header}
\end{figure*}

\subsection{App Dataset}
\label{App-Dataset}

For our evaluation we downloaded a set of popular free apps from the US Google Play store in August 2023.
Following previous work~\cite{scoccia_leave_2020,shuba_nomoats_2020,han_price_2020}, we identified the apps based on Google's lists of popular apps in 49 app categories on the Google Play store.
Each list contained 45 apps.
The 49 app categories include 17 game categories~\cite{GooglePlayTopAppCategories}.
The sets of apps in the different categories are disjoint as each app can be assigned one category on the Google Play store~\cite{GooglePlayTopAppCategories}.
The 45 apps listed for each of the 49 categories are ``heavily influenced by popularity''~\cite{GooglePlayTopAppCharts}.\footnote{Apps in the Communication and Productivity categories had the most installs while those in the Events category had the least. Appendix Figure~\ref{fig:app-installs} shows the app install ranges for the apps in all 49 app categories.}
We selected popular free apps for our study because those are most likely to rely on an ad-based business model that implies selling personal information or sharing such with third parties (Section~\ref{The-CCPA's-Opt-out-Right}).
Using Raccoon~\cite{OnyxbitsRaccoon}, we successfully downloaded 1,896/2,205 (86\%) of the listed apps.
The remaining apps were incompatible with our test device, removed from the Google Play store before we could download them, or not successfully downloaded.
We made two download attempts per app.
For our GPC opt-out evaluation we successfully performed a dynamic analysis of 1,811/1,896 (96\%) apps resulting in our \textit{app dataset}.
For the remaining apps we did not collect any or only little network traffic, for example, due to apps' crashing during the analysis.

\subsection{Measuring Opt-out Effects}

We measure the effects of opting out---i.e., whether an app stopped selling or sharing personal information---in three ways: we measure whether an app (1) sets CCPA privacy flags to opt out (such as the US Privacy String~\cite{IABUSPrivacyString}), makes fewer connections to ad tracking domains, or (3) discloses fewer device identifiers to third parties. 
As to the first measure, it is the responsibility of ad networks to provide functionality in their ad libraries for app operators to identify ad requests to which the CCPA opt-out right applies.
If an app operator enables such functionality, a privacy flag should be set on the user's device and possibly also be attached to outgoing ad requests to third-party ad networks, as shown in Figure~\ref{fig:GPC_Android_Header}.
Thus, as privacy flags are generally communicated client-to-server, for example, ad networks retrieve the US Privacy String from an Android's device local \texttt{SharedPreferences}~\cite{IABUSPrivacyStringStorage}, we can use client-side analysis to determine the opt-out status.
If a third party receives an ad request with a privacy flag, it knows to discard personal information upon receipt and serve an ad not based on sold or shared personal information.

Beyond changing privacy flag values, a successful opt-out can manifest itself through other observable changes in an app's behavior.
An app can comply by not making any ad calls. 
Thus, monitoring an app's connections to third-party ad tracking domains serves as another indicator for opt-out adherence.
Further, an app may continue to connect to third-party ad tracking domains but stop the sharing of identifying data. 
Thus, we also examine the impact of GPC signals on data sharing practices, specifically, the disclosure of device identifiers. 
A reduction in the sharing of such identifiers can indicate that an opt-out request has been respected since a third party is now more limited in associating particular data with a particular user or device.

Android's ad ecosystem is integrated into the broader online ad ecosystem and relies on the delivery of ads via the web protocol.
An opt-out request, and the corresponding setting of a privacy flag, can be initiated in several ways.
Figure~\ref{fig:GPC_Android_Header} shows how a user initiates an opt-out request by sending a GPC header.
However, it is likely that many apps do not yet have functionality to process GPC signals.
Especially, since Android does not have a platform-level GPC setting, apps operators may not have set up server-side code to detect GPC signals. 
Thus, we also evaluate to which extent privacy flags are set via in-app opt-out interfaces or disabling the AdID.
The lack of platform-level support for GPC, however, does not principally change the legal status of GPC as a mandatory opt-out setting on Android and, more generally, any web-based platform.

\subsection{UI Opt-out Evaluation}
\label{UI-Opt-out-Evaluation}

In our app-level UI opt-out evaluation, for determining whether the CCPA opt-out right is applicable to an app we relied on app operators' respective declarations in their privacy policies.
We first chose from our app dataset 100 random apps whose operators notify California residents in their privacy policies that they have a CCPA opt-out right. 
To ensure that a privacy policy is applicable to an analyzed app we relied on the link from each app's US Google Play store Data Safety page to the policy.
If necessary, we further followed links from a policy to US state- or California-specific policies.
Similar to previous work~\cite{kollnig2021_consent_analysis}, we categorize each app's opt-out setting as follows:

\begin{itemize}
    \setlength{\itemsep}{1pt}
    \setlength{\parskip}{0pt}
    \setlength{\parsep}{0pt}
    \item \textbf{Opt-out Right Setting}: The app has a dedicated CCPA opt-out setting, e.g., toggle or button, to opt out of the selling or sharing of personal information per the CCPA.
    \item \textbf{Generic/SDK/Cookie Opt-out Setting}: The app has an opt-out setting without mentioning that it relates to the selling or sharing of personal information or that users have a right to opt out. This category covers generic opt-out settings as well as settings for cookie and SDK opt-outs.
    \item \textbf{Opt-out Right Form}: The app has a CCPA right form, e.g., requiring California residents to enter their email address, to opt out of the selling or sharing of personal information or claim other CCPA rights.
    \item \textbf{Privacy Policy}: The app only has a privacy policy link but no opt-out setting inside the app. We do not count opt-out settings that are only accessible via a privacy policy. If there is just a link to a ``privacy policy'' inside the app, which then links to an opt-out setting, it would not be ``clear and conspicuous'' and insufficient per the CCPA if the app does not respect opt-out signals via GPC.
    \item \textbf{No Opt-out Setting or Privacy Policy}: The app is missing both an opt-out setting and a privacy policy link.
\end{itemize}

We evaluated whether an app has an opt-out setting, for example, under a ``privacy settings'' or ``settings'' menu. 
If clicking on such a menu item took us to an external website, we followed the respective procedure there.
For a subset of 38 apps---those with opt-out right setting or generic/SDK/cookie opt-out setting---we performed a dynamic analysis.
We did not evaluate apps with opt-out right form because submitting such form would not have immediate impact visible in the apps' network traffic.
For 31/38 apps we were able to capture the network traffic in our two test conditions.
For the remaining apps we did not collect any or only little network traffic, for example, due to apps' crashing during the analysis.

The two test conditions under which we evaluated each app are the following:

\begin{enumerate}[1.]
    \setlength{\itemsep}{1pt}
    \setlength{\parskip}{0pt}
    \setlength{\parsep}{0pt}
    \item \textbf{Not Opted Out}: Use of the main functionality of the app (e.g., if the app is a game, play it) with the app's opt-out setting, if any, set to not opted out. This condition is the base condition under which we expect an app to sell and share California residents' personal information.
    \item \textbf{Opted Out}: Use of the main functionality of the app with the app's opt-out setting, if any, set to opted out.
\end{enumerate}

\begin{figure}[t!]
    \centering
    \includegraphics[width=0.47\textwidth]{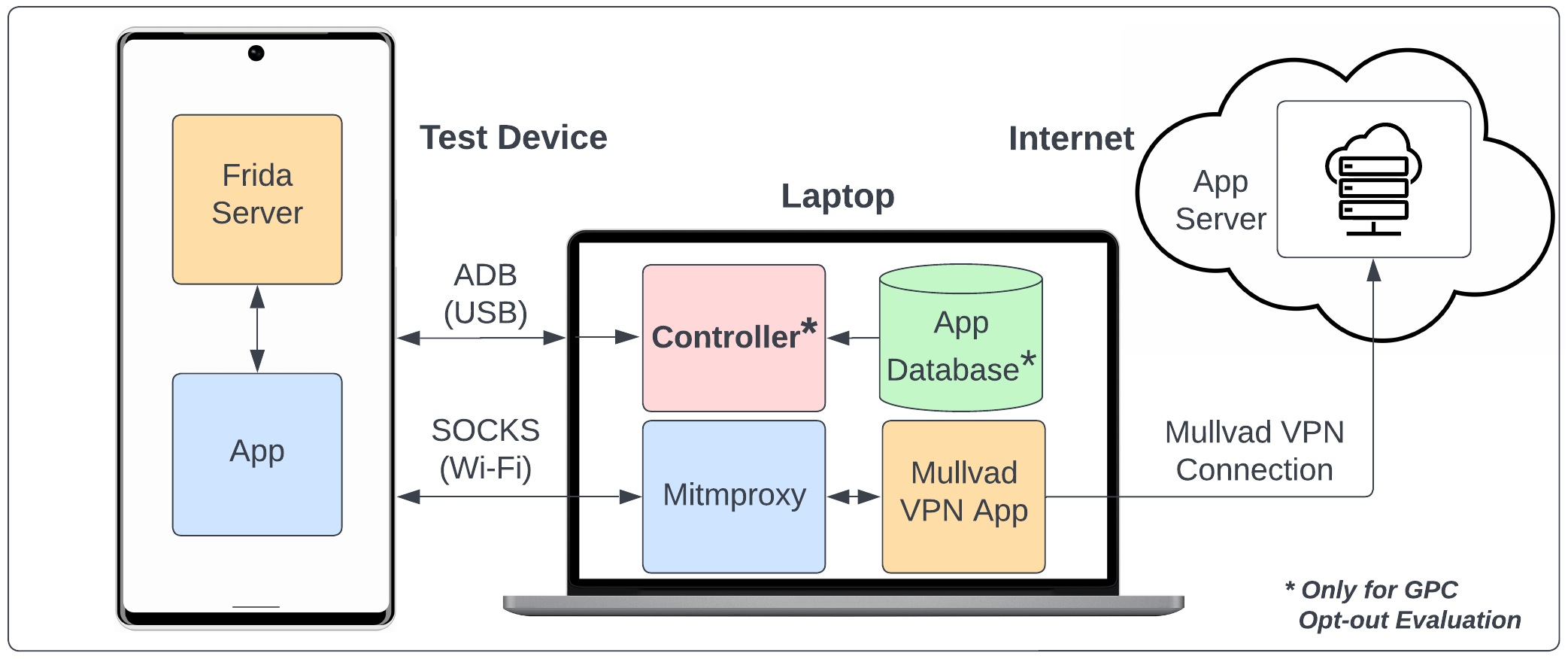}
    \caption{Overview of our dynamic analysis setup for capturing app network traffic. Appendix~\ref{Dynamic-Analysis-Details} contains a comprehensive description of our dynamic analysis for the UI and GPC opt-out evaluation.}
    \Description{Overview of our dynamic analysis setup for capturing app network traffic.}
    \label{fig:analysis-setup}
\end{figure}

During our evaluation we performed for each app a manual dynamic analysis.
For each condition, we captured each app's network traffic for 30 seconds, similar to previous work~\cite{kollnig2021_consent_analysis}, using a Google Pixel 6a with Android 13, our \textit{test device}.
We found in preliminary tests that this time period is sufficient to understand the app's ad tracking behavior and that the order of the conditions does not impact the results. 
If necessary to use the app, we created an account and logged into the app before evaluating the two test conditions.
For each app, when switching conditions, we cleared the persistent storage and \texttt{SharedPreferences} of the test device.
Upon an app's request, we granted all permissions except for location permission as an additional measure to prevent apps from detecting our real location as we were not based in California.

For the analysis of network traffic, we followed the state-of-the art in the literature: use of Mitmproxy~\cite{Mitmproxy} and circumvention of certificate pinning with Frida~\cite{Frida}.
For capturing an app's network traffic, we connected the test device to a laptop using Mitmproxy in SOCKS mode~\cite{Mitmproxy}.
To issue commands to the test device, e.g., to start Mitmproxy, we connected the test device to the laptop via the Android Debug Bridge (ADB)~\cite{AndroidADB}.
We obtained a California IP address via a Mullvad VPN server~\cite{Mullvad} to ensure applicability of the CCPA. %
During our analysis we also manually took screenshots of the opt-out UI, if any.
We performed out analysis in July and August 2025.
Figure~\ref{fig:analysis-setup} shows an overview of our analysis setup.

\subsection{GPC Opt-out Evaluation}
\label{GPC-and-AdID-Opt-Out-Evaluation}

In addition to our UI opt-out evaluation, we performed a GPC opt-out evaluation.
The ``selling'' or ``sharing'' of personal information by an app operator corresponds to a third party's ``buying'' or ``collecting'' of personal information.
Thus, to determine whether an app is selling or sharing personal information we identified the most prevalent third-party tracking domains across all apps in our app dataset.
Our starting point for this task was the X-Ray Tracker List~\cite{10.1145/3176246,xray-blacklist}.
We found that a set of 29 third-party tracking domains from the X-Ray Tracker List lead to a 99\% coverage with at least one tracking domain---whose privacy policy allowed to (1) buy or collect or (2) sell or share personal information per the CCPA---in each app in our app dataset (Section~\ref{CCPA-Opt-Out-Right-Applicability}).\footnote{We combined trackers with different subdomains in our app dataset into one domain to match the tracker domains in the X-Ray Tracker List. For example, we combined \url{app.adjust.com}, \url{view.adjust.com}, \url{app.us.adjust.com}, and \url{s2s.adjust.com} into \url{adjust.com}.}
For the privacy policy evaluation we included developer documentation and looked for explicit references that---per the CCPA---allows the buying or collecting of personal information from app operators or the selling or sharing of such.

To illustrate, Meta's policy~\cite{MetaStateSpecificPolicy} indicates that it is buying or collecting personal information per the CCPA via its libraries as it advises app operators integrating those that they may ``share with Meta Personal Information about California residents which would otherwise constitute a Sale or Sharing [...].''
In another example, per InMobi's privacy policy~\cite{InMobiPrivacyPolicy}, ``InMobi may be deemed to be `selling' personal information, [...]'' indicating InMobi's further downstream sale of California residents' personal information after buying or collecting such from the app operators who integrate its libraries.
We performed our evaluation conservatively. 
If a policy stated that a third party does not share or sell personal information per the CCPA, we accepted such statement as true.
Our methodology could be applied to cover cases where the operator claims not to sell or share personal information but does so in practice, for example, as it appears from the operator's app network traffic, possibly in combination with a privacy flag not set to opt-out.
In such case the operator needs to either adapt the claim or cease the practice.

To evaluate the likelihood to which an app in our app dataset is part of a business, we selected a random sample of 100 apps from our app dataset for determining whether their operators self-declared to run a ``business'' in their privacy policies.
We performed this analysis on the 83 sample apps in our app dataset that had a Google Play Data Safety page with a link to their privacy policy and were still available at the time of the sampling.
Given that an app's Data Safety page linked to a privacy policy, we assumed that the linked policy applies to the app.

The setup and procedure for our GPC opt-out evaluation is similar to our UI opt-out evaluation, though, automated.
After downloading the apps, we ran an automated dynamic analysis for each app individually proceeding in batches for a total of two weeks with the analysis of one app taking about 5 minutes.
We performed this analysis for the apps in our app dataset in October 2023.
As shown in Figure~\ref{fig:analysis-setup}, we instrumented the test device with a shell script, the \textit{controller}, which ran on our laptop.
The controller fetched an app from the app database on the laptop, installed it on the test device, ran it in the foreground under four test conditions, and uninstalled it.
If an app update was triggered right after installing an app, it would not have been possible to run the app immediately afterwards, which we checked for.
\footnote{Appendix~\ref{Dynamic-Analysis-Details} contains further details of our dynamic analysis with Appendix Figure~\ref{fig:analysis_procedure} showing the procedure in detail.}

We wanted to find out if California residents can exercise their opt-out rights by sending GPC signals.
To which extent do Android apps respect GPC signals despite the lack of platform-level API support for such?
We were also interested in the effect of disabling the AdID.
Thus, we tested each app under four conditions encompassing all possible combinations of access to the AdID (yes/no) and sending GPC signals (yes/no). 
Similar to previous work~\cite{kollnig2021_consent_analysis}, for each condition, we captured each app's network traffic for 30 seconds.
We also added a timeout of 30 seconds before running an app's first condition so that the app is fully initialized before capturing of the network traffic begins~\cite{nguyen_freely_2022}.
The four test conditions under which we evaluated each app in our app dataset are the following:

\begin{enumerate}[1.]
    \setlength{\itemsep}{1pt}
    \setlength{\parskip}{0pt}
    \setlength{\parsep}{0pt}
    \item \textbf{AdID}: The app has access to the AdID. This condition is the base condition under which we expect some apps to sell and share California residents' personal information.
    \item \textbf{AdID + GPC}: The app has access to the AdID, but all HTTP requests are sent with a GPC header. The app operator is prohibited from selling and sharing personal information per the CCPA. The app should notify integrated ad networks of the opt-out via respective privacy flags.
    \item \textbf{No AdID}: The app does not have access to the AdID. Google prohibits targeting users with personalized advertising~\cite{PlayConsoleHelpAds}. If the app recognizes the AdID setting as a valid opt-out preference per the CCPA, it should notify integrated ad networks of the opt-out via respective privacy flags.
    \item \textbf{No AdID + GPC}: The app does not have access to the AdID, and all HTTP requests are sent with a GPC header. This condition evaluates the cumulative effect of no AdID access and GPC. The app should notify integrated ad networks of the opt-out via respective privacy flags.
\end{enumerate}

\begin{figure}[t]
    \centering
    \includegraphics[width=0.48\textwidth]{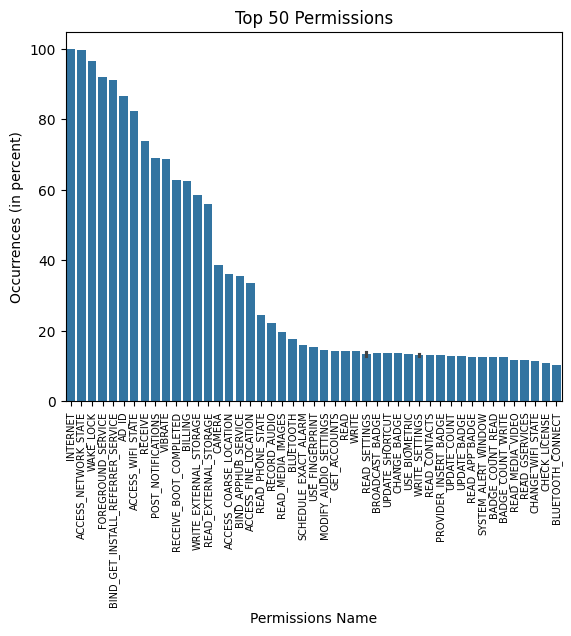}
    \caption{The top 50 permissions declared by apps in our app dataset ($n=1,811$) in their manifest files.}
    \Description{The top 50 permissions declared by apps in our app dataset ($n=1,811$) in their manifest files.}
    \label{fig:permissions}
\end{figure}

In the AdID + GPC and No AdID + GPC conditions we injected GPC headers into all requests, for both HTTP and HTTPS, using Mitmproxy.
In the AdID and No AdID conditions, we injected dummy headers into all requests to minimize any differences that might occur due to header injection in the other two conditions. 
In all four conditions the controller granted all permissions requested by an analyzed app at install time.\footnote{Different from the UI Opt-out evaluation (Section~\ref{UI-Opt-out-Evaluation}), for the GPC opt-out evaluation we opted to grant apps location permission as we also wanted to explore location tracking. Appendix Figure~\ref{fig:permissions} shows the top 50 permissions that the 1,811 apps in our app dataset declared in their manifest file.}
We tested whether granting or denying permissions made a difference for apps' ad tracking behavior, which was not the case.
When switching conditions, our controller cleared the persistent storage and \texttt{SharedPreferences} of the test device.

\subsection{Limitations}

Our study has various limitations.
First, our evaluation of the applicability of the CCPA opt-out right is based on privacy policies that app operators link from their app's Play store Data Safety page.
It may be that a policy is not applicable to an app or a policy does not distinguish between multiple apps or the operator's website, each of which may have different data practices.
Further, by its very nature, dynamic analysis only runs code partially leaving us with an incomplete picture of apps' behavior. 
Also, for our GPC opt-out evaluation we evaluated apps without user interaction save for automatically enabling and disabling the AdID setting.\footnote{Automatically enabling or disabling the AdID may be impacted by pop-up screens or other interruptions. We encountered some apps with unexpected AdID access. We manually re-analyzed these apps and confirmed the results (Section~\ref{Disclosure-of-Device-Identifiers}).}
For determining opt-out effects we evaluated privacy flags, domain connections, and sharing of device identifiers.
However, those are only proxies.
For example, an ad network may simply disregard a user's right despite a privacy flag being set to opt-out.
Our methodology does not allow us to observe server-to-server data sharing, for example, of device identifiers.
Further, some apps we evaluated likely used encryption with certificate pinning. 
While we followed state-of-the-art methods for certificate-unpinning~\cite{xiao_lalaine_2023,Kollnig_2022,nguyen_freely_2022}, we were likely not able to decrypt all encrypted traffic, which is an inherent limitation of any dynamic analysis of third-party code.
For our identification of ad tracking domains we did not perform an independent evaluation and relied on previous research that led to creation of the X-Ray Tracker List~\cite{10.1145/3176246,xray-blacklist}.
Similar to previous work~\cite{277218}, we used a commercial VPN, Mullvad, to simulate a California location so that the CCPA is applicable. 
Apps may have identified our real location, e.g., via Mullvad's IP address or our time zone setting, and behaved differently.
Thus, our results may not be an accurate reflection of app behavior for California residents.
However, we are not aware of such behavior.
Notably, in our GPC opt-out evaluation we did not notice different behavior in apps with and without GPS access.

\section{UI Opt-out Evaluation}
\label{UI-Opt-Out-Evaluation}

\textbf{Main Result}: Only 48 of the 100 apps we evaluated implement the legally mandated CCPA opt-out setting, which suggests that CCPA opt-out right compliance on the Android platform is generally low.

\subsection{CCPA Opt-out Right Applicability}

For all 100 apps for which we performed our UI opt-out evaluation we ensured that the CCPA opt-out right is applicable.
The privacy policies of all 100 apps included respective statements.

\subsection{Opt-out Effects}
\label{UI-Opt-Out-Effects}

\begin{figure}[t!]
    \centering
    \includegraphics[width=0.223\textwidth]{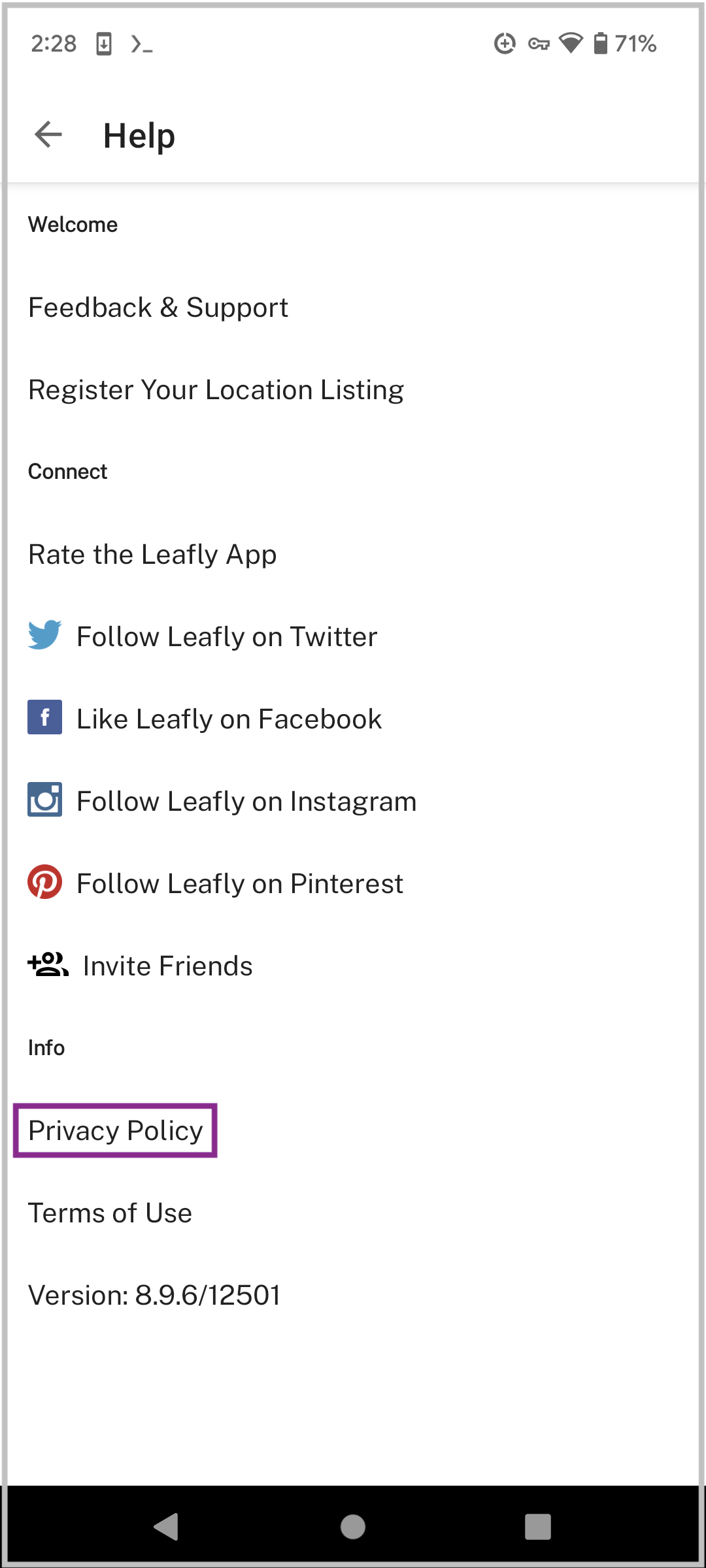}
    \hspace{0.1in}
    \vspace{0.1in}
    \includegraphics[width=0.223\textwidth]{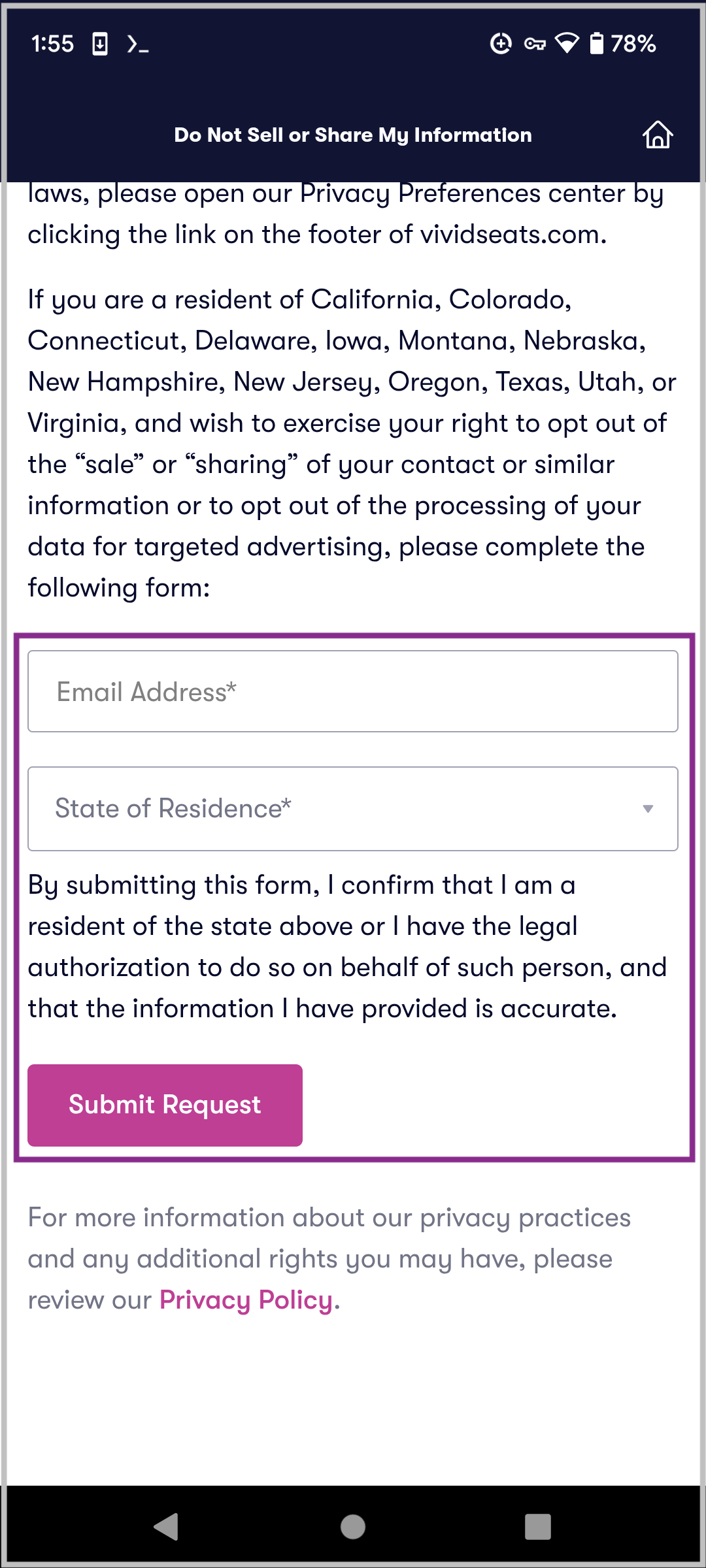}
    \hspace{0.1in}
    \includegraphics[width=0.47\textwidth]{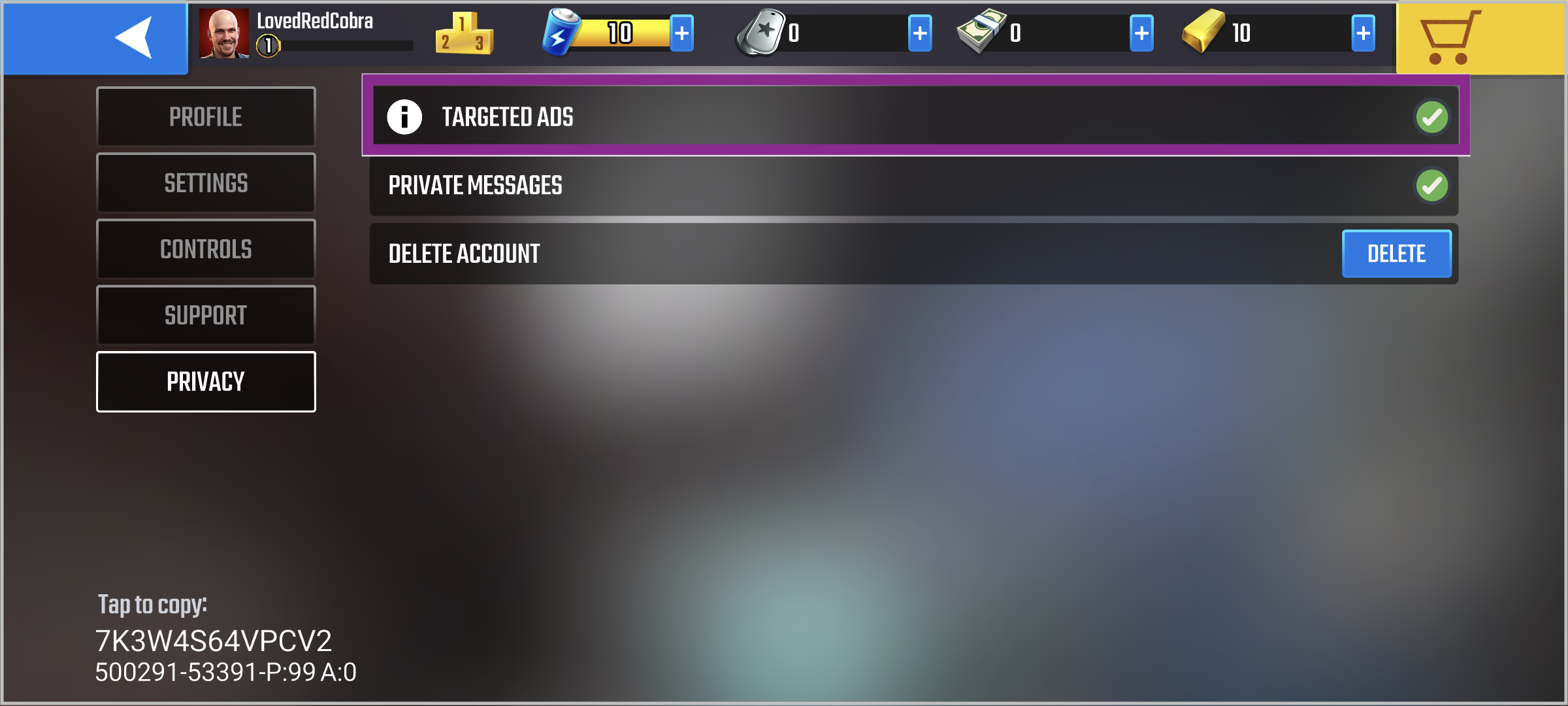}
    \caption{Example UIs of three apps: \textit{Leafly: Find Cannabis and CBD}~\cite{Leafly} with just a privacy policy link (top left), \textit{Vivid Seats | Event Tickets}~\cite{VividSeats} with an opt-out right form (top right), and \textit{Pure Sniper: Gun Shooter Games}~\cite{PureSniper} with a generic/SDK/cookie opt-out setting (bottom). Figure~\ref{fig:opt-out-right-screenshot} shows an example UI of a dedicated opt-out right setting.}
    \Description{Example UIs of three apps.}
    \label{fig:opt-out-uis}
\end{figure}

\begin{table}[t]
    \centering
    \resizebox{\columnwidth}{!}{
    \begin{tabular}{lccc}
        \toprule
        & \multicolumn{1}{c}{\textbf{Count}} & \multicolumn{1}{c}{\textbf{Confidence Interval}} \\
        \midrule
        \textbf{Opt-out Right Setting} & 30 & 21\%--39\% \\
        \textbf{Generic/SDK/Cookie Opt-Out Setting} & 8 & 3\%--13\% \\
        \textbf{Opt-out Right Form} & 10 & 4\%--16\% \\
        \textbf{Privacy Policy Link} & 45 & 35\%--55\% \\
        \textbf{No Opt-out Setting or Privacy Policy Link} & 7 & 2\%--12\% \\
        \bottomrule
    \end{tabular}
    }
    \hspace{0.2in}
    \caption{Counts of opt-out settings for the 100 apps for which we performed the UI opt-out evaluation and confidence intervals for the prevalence of respective settings in our app dataset. Figure~\ref{fig:opt-out-uis} shows example UIs.}
    \label{tab:ui-opt-out-statistics}
\end{table}

As Table~\ref{tab:ui-opt-out-statistics} shows, only 48 out of 100 apps implement the legally mandated CCPA opt-out setting.
The remaining 52 apps either have just a privacy policy link (45) or no opt-out setting or privacy policy link (7).
As an example, calculating the confidence interval for the proportion of 45/100 (45\%) apps with just a privacy policy link, using z-scores, indicates that with 95\% confidence the true proportion of apps in our app dataset with only such link is 35\%--55\%.
Thus, even without evaluating any network traffic, our results suggest the broad existence of non-compliance.
While apps in some game categories tend to have higher percentages of non-compliant apps, non-compliance is generally observable across all app categories (Appendix Table~\ref{tab:ui-opt-out-compliance-statistics}).

Turning to our network traffic analysis, some apps did not reflect the opted out state in their privacy flags.
Thus, even if California residents opt out, those apps are still selling and sharing personal information.
Out of the 14 apps that included the third-party ad network Vungle~\cite{Vungle}, two apps still set the \texttt{ccpa status} flag to \texttt{opted\_in} in the opted out condition so that Vungle continued to collect and buy personal information.
Similarly, one app also still had Google's Restricted Data Processing (RDP)~\cite{GoogleRDP} set to \texttt{0}.
These results show that opt-out rights are not always respected as app operators do not notify third parties of users' opt-outs so that those remain opted in.
We did not see differences for connecting to ad tracking domains and disclosing device identifiers between opted out and not opted out conditions.

\section{GPC Opt-out Evaluation}
\label{GPC-Opt-Out-Evaluation}

\textbf{Main Result}: While we estimate with 95\% confidence that 62\%--81\% of apps in our app dataset must respect the CCPA opt-out right, many apps do not do so.
For example, when sending GPC signals and disabling apps' access to the AdID, 338 apps still had the \texttt{ccpa status} of the ad network Vungle set to \texttt{opted\_in} while only 26 had set it to \texttt{opted\_out}.

\subsection{CCPA Opt-out Right Applicability}
\label{CCPA-Opt-Out-Right-Applicability}

\begin{figure}[t!]
    \centering
    \includegraphics[width=3.2in]{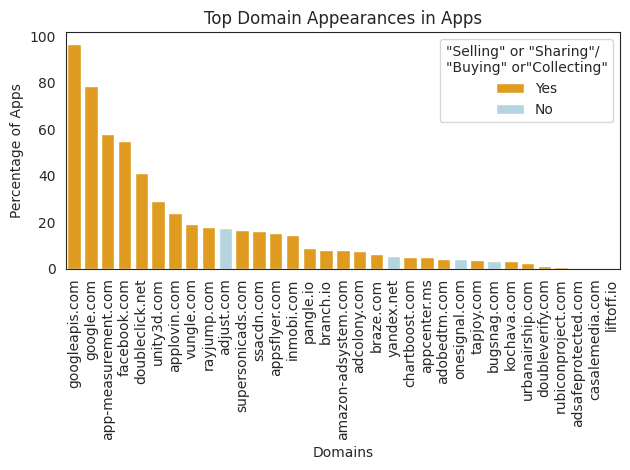}
    \caption{For 29/33 (88\%) of the most prevalent ad tracking domains in our app dataset their privacy policies allow the ``Selling'' or ``Sharing''/``Buying'' or ``Collecting'' of personal information per the CCPA.}
    \Description{For 29/33 (88\%) of the most prevalent ad tracking domains in our app dataset their privacy policies allow the ``Selling'' or ``Sharing''/``Buying'' or ``Collecting'' of personal information per the CCPA.}
    \label{fig:ad-networks}
\end{figure}

\begin{figure}[t!]
   \centering
   \includegraphics[width=3.2in]{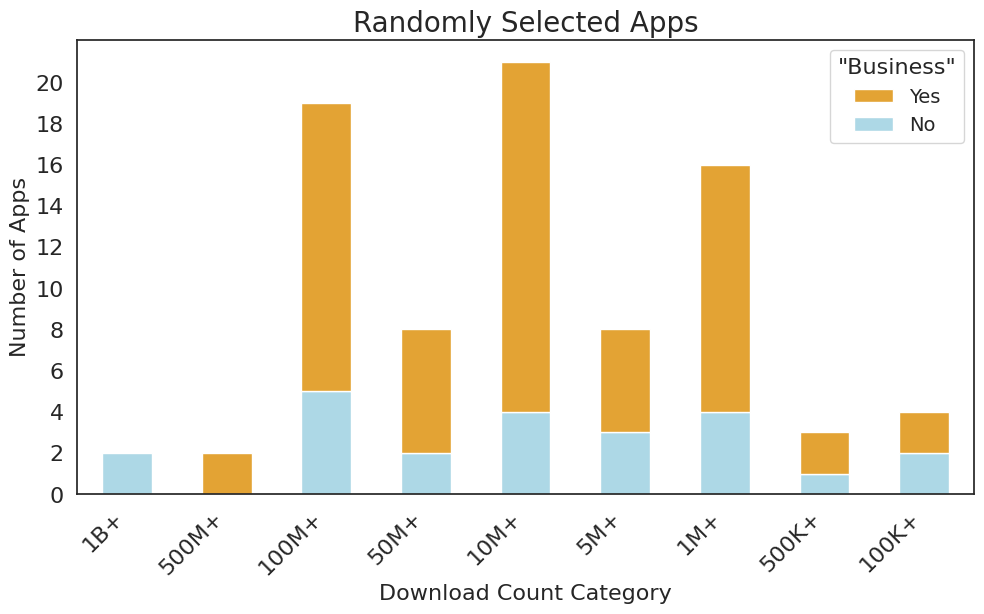}
   \caption{For 60/83 (72\%) randomly sampled apps from our app dataset their privacy policies stated that the operators run a ``business.''}
   \Description{For 60/83 (72\%) randomly sampled apps from our app dataset their privacy policies stated that the operators run a ``business.''}
   \label{fig:app-installs-count}
\end{figure}

Evaluating how many apps are selling or sharing personal information, we find that 1,787/1,811 (99\%) of the apps in our app dataset connect to at least one of the most prevalent 29 ad tracking domains with a privacy policy stating to (1) buy or collect or (2) sell or share personal information per the CCPA (Figure~\ref{fig:ad-networks}).\footnote{Performing the evaluation for all third-party tracking domains on the X-Ray Tracker List only increases the coverage by two apps to 1,789/1,811 (99\%), which highlights the concentration of companies in the online ad industry.}
Further, for the random sample of 83 apps (Section~\ref{GPC-and-AdID-Opt-Out-Evaluation}), 60/83 (72\%) of their policies 
implied that the app operators run a business per the CCPA
by mentioning the applicability of the CCPA or the explicit selling or sharing of personal information (Figure~\ref{fig:app-installs-count}).
Thus, we estimate that $1,787*0.72=1,287$ apps in our app dataset are subject to the CCPA opt-out right.
Calculating the confidence interval for the proportion of 60/83 (72\%) sample apps, using z-scores, indicates that with 95\% confidence the true proportion of app operators running a business is 63\%--82\%.
Applying this range to the 1,787 apps that connect to at least one of the 29 most prevalent tracking domains, we arrive at an estimate of 1,126 to 1,465 (62\%--81\%) apps in our app dataset being subject to the CCPA opt-out right.

\subsection{Opt-out Effects}
\label{opt-out-effectiveness}

Overall, we find little evidence that the apps in our app dataset respect GPC.
Notably, we observe no substantial impact of GPC on apps' notifications to third parties via the IAB's US Privacy String~\cite{IABUSPrivacyString} and company-specific privacy flags (Section \ref{Setting-of-CCPA-Privacy-Flags}). 
We also do not observe a decrease in the number of connections to ad tracking domains (Section \ref{Connection-to-Ad-Network-Domains}) or the disclosure of device identifiers (Section \ref{Disclosure-of-Device-Identifiers}).

\subsubsection{Setting CCPA Privacy Flags}
\label{Setting-of-CCPA-Privacy-Flags}

The IAB provides an industry-wide CCPA privacy flag in the form of the US Privacy String~\cite{IABUSPrivacyString}, to be implemented as an HTTP cookie or a JavaScript property.\footnote{On January 31, 2024 the US Privacy String was deprecated and integrated into the IAB's Global Privacy Platform (GPP) that covers privacy flags for multiple US states and international jurisdictions~\cite{GPP}. The IAB now recommends use of GPP instead of the US Privacy String. GPP contains the \texttt{usca} California privacy section corresponding to the US Privacy String~\cite{GPPCalifornia}. Since we performed our analysis before the deprecation date we focus here on the US Privacy String. 
Both the US Privacy String as well as its GPP equivalent can be stored on Android devices using \texttt{SharedPreferences} for the app context~\cite{IABUSPrivacyStringStorage,GPP}.
} 
We find that when opting out via GPC, privacy flags generally do not change to reflect the CCPA opt-out status.
However, some apps opt out users regardless of whether they receive GPC signals or have access to the user's AdID.

\begin{figure}[t!]
    \centering
    \includegraphics[width=\columnwidth]{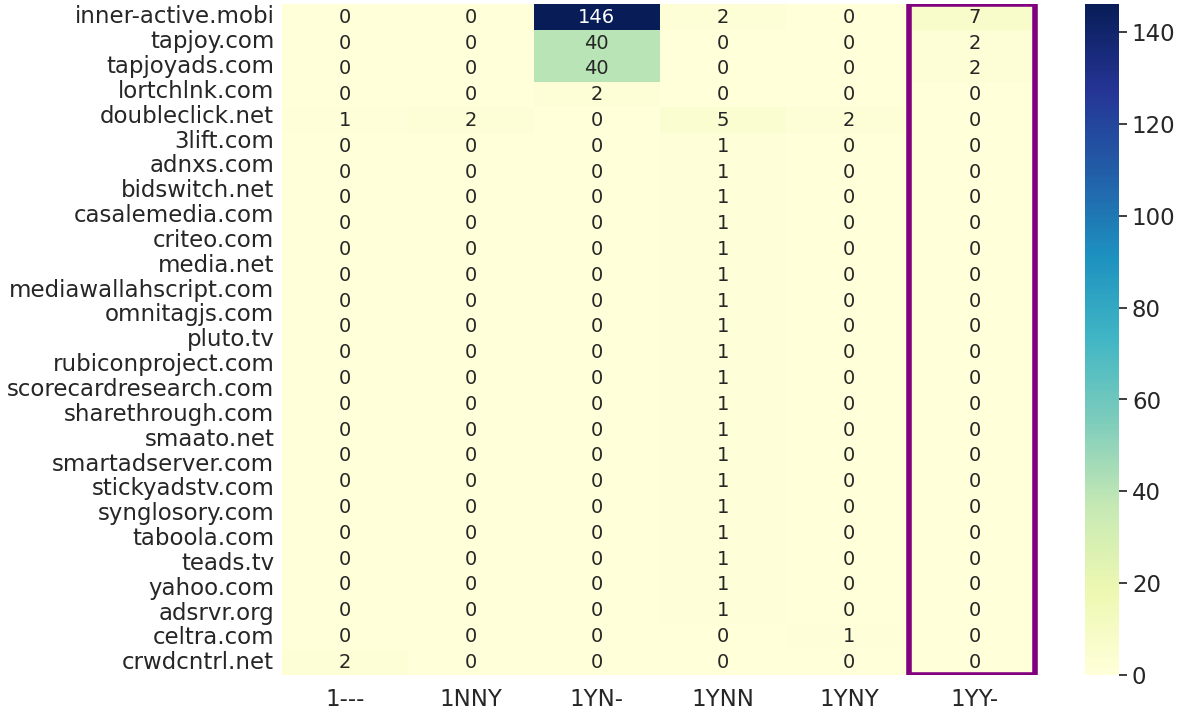}
    \includegraphics[width=\columnwidth]{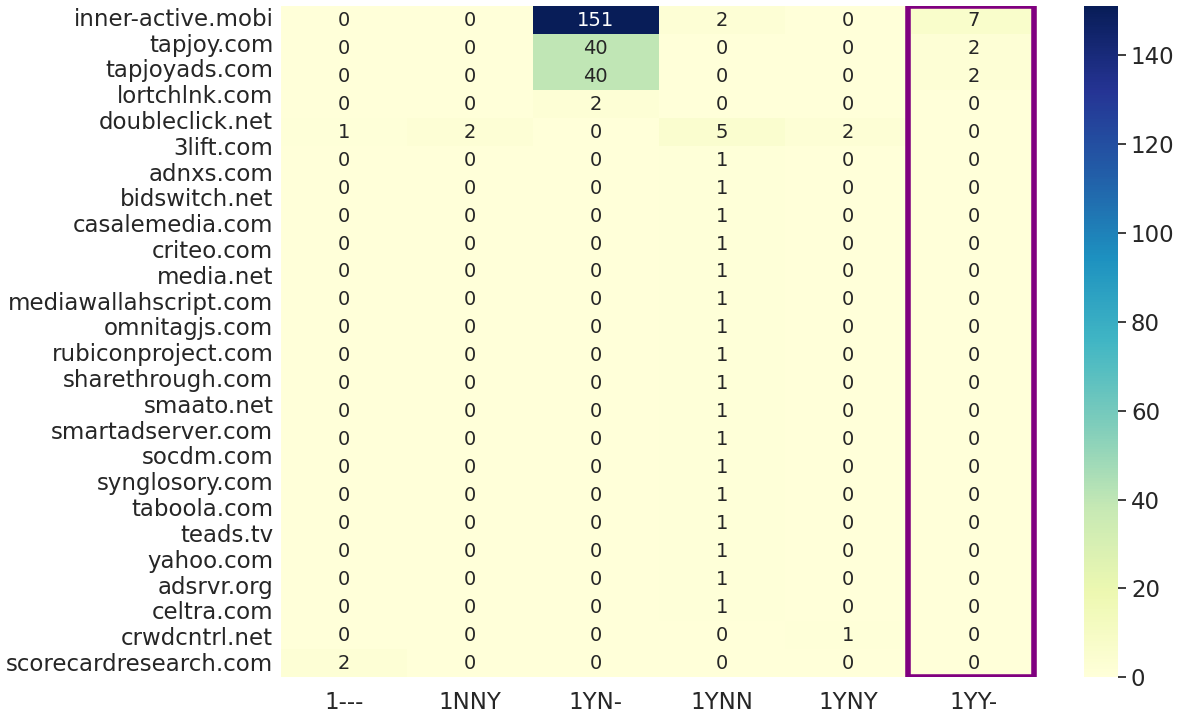}
    \caption{Counts of US Privacy String values in the No AdID (top) and No AdID + GPC (bottom) conditions. The rightmost columns show the opt-out counts of US Privacy String values that apps in our app dataset sent to third-party domains. For example, in the No AdID condition only 7/155 apps sent a \texttt{1YY-} value to \url{inner-active.mobi} indicating the opt-out status while the remaining apps sent \texttt{1YN-} (146/155) and \texttt{1YNN} (2/155) values indicating the opt-in status. These results did not substantially change when enabling GPC in the No AdID + GPC condition. Notably, the number of apps sending a \texttt{1YY-} value stayed the same (7/160). The counts are per app and mutually exclusive for a given domain but not across domains as one app can connect to multiple domains. We only include domains with at least one US Privacy String value in our network traffic.}
    \Description{Counts of US Privacy String values in the No AdID (top) and No AdID + GPC (bottom) conditions.}
    \label{fig:us-privacy-string_no_adid_no_gpc}
\end{figure}

\paragraph{US Privacy String.}

\begin{figure}[t!]
    \centering
    \includegraphics[width=\columnwidth]{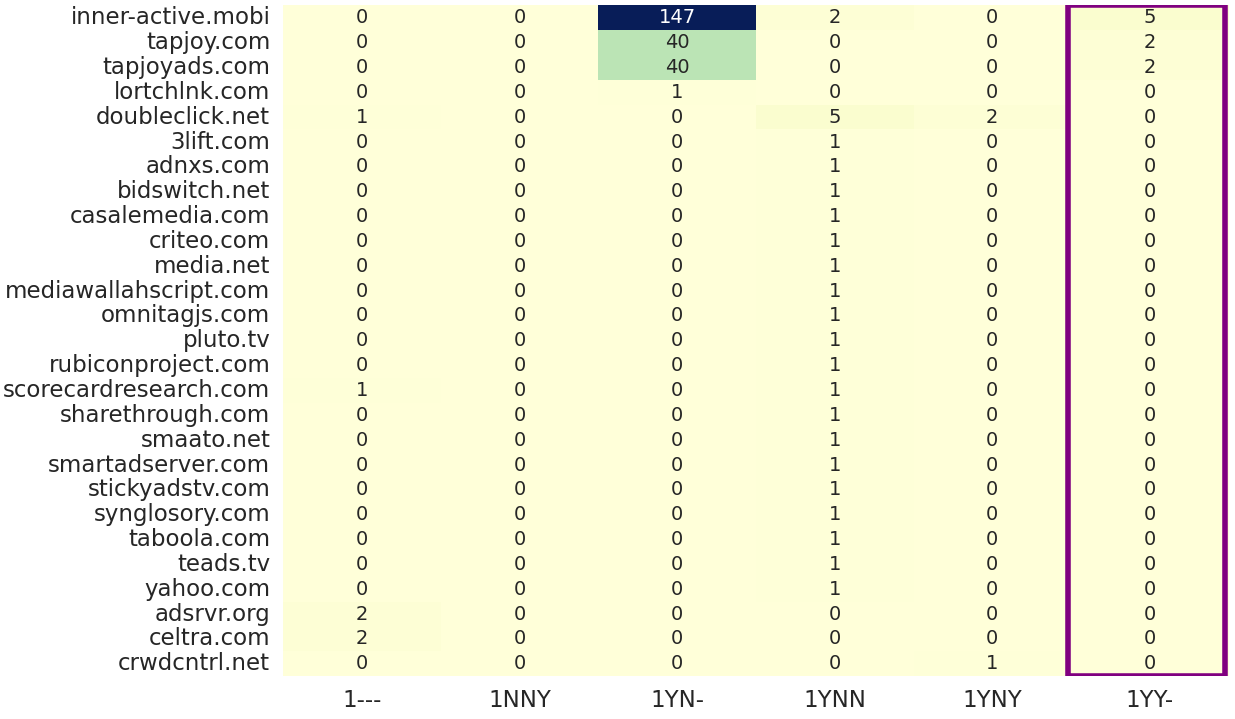}
    \includegraphics[width=\columnwidth]{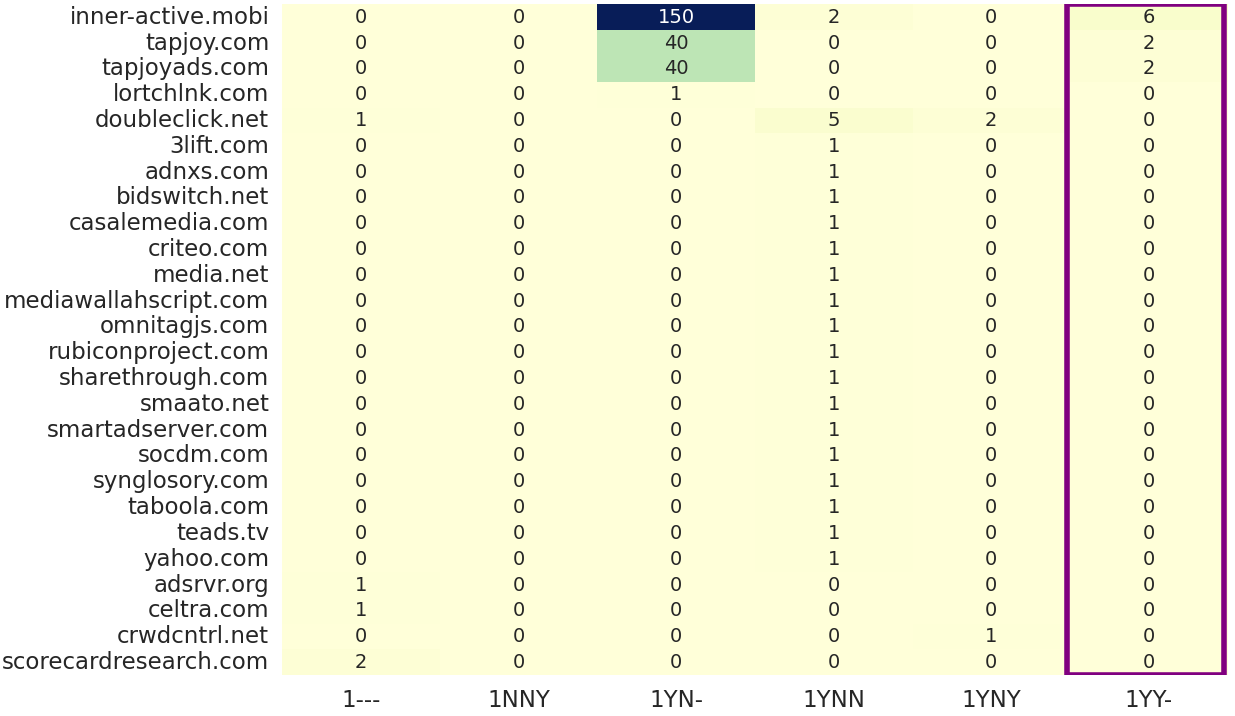}
    \caption{Counts of US Privacy String values in the AdID (top) and AdID + GPC (bottom) conditions.}
    \Description{Counts of US Privacy String values in the AdID (top) and AdID + GPC (bottom) conditions.}
    \label{fig:us-privacy-string_adid_no_gpc}
\end{figure}

For 199/1,811 (11\%) apps in our app dataset we found a US Privacy String key, \texttt{us\_privacy}, and a corresponding US Privacy String value in their network traffic.
However, the value of the US Privacy String was rarely set to opt-out, that is, rarely had a \texttt{Y} in its third position, which would indicate a ``Yes'' for a California resident's status of being opted out~\cite{IABUSPrivacyString} (Figure~\ref{fig:us-privacy-string_no_adid_no_gpc}). 
Also, disabling the AdID did not lead to substantially different opt-out rates as the comparison between the AdID and No AdID conditions shows (Appendix Figure~\ref{fig:us-privacy-string_adid_no_gpc}).
Overall, the rate of \texttt{1YY-} US Privacy String values, which indicate that a California resident's opt-out right is respected, is low regardless of the tested condition.
Our results suggest that turning on GPC does not help California residents to exercise their CCPA opt-out right as measured via the US Privacy String.
Disabling the AdID did not change this result.

\begin{table}[t]
    \centering
    \resizebox{\columnwidth}{!}{
    \begin{tabular}{lcccc}
        \toprule
        & \multicolumn{1}{c}{\textbf{Google}} & \multicolumn{1}{c}{\textbf{Supersonic/Unity}} & \multicolumn{1}{c}{\textbf{Vungle}} \\
        & \multicolumn{1}{c}{\textbf{\texttt{rdp}}} & \multicolumn{1}{c}{\textbf{\texttt{do\_not\_sell}}} & \multicolumn{1}{c}{\textbf{\texttt{ccpa status}}} \\
        & \multicolumn{1}{c}{\texttt{0, 1}, \%} & \multicolumn{1}{c}{\texttt{false, true}, \%} & \multicolumn{1}{c}{\texttt{opted\_in, opted\_out}, \%} \\
        \midrule
        \textbf{AdID} & 12, 29, 71\% & 152, 2, 1\% & 339, 23, 6\% \\
        \textbf{AdID + GPC} & 12, 30, 71\% & 154, 2, 1\% & 341, 23, 6\% \\
        \textbf{No AdID} & 12, 31, 72\% & 156, 2, 1\% & 337, 24, 7\% \\
        \textbf{No AdID + GPC} & 12, 31, 72\% & 155, 4, 3\% & \textbf{338, 26, 7\%} \\
        \bottomrule
    \end{tabular}
    }
    \hspace{0.2in}
    \caption{Counts and percentages of apps that set company-specific privacy flags. For example, when sending GPC signals and disabling apps' access to the AdID, 338 apps still had the \texttt{ccpa status} of the ad network Vungle set to \texttt{opted\_in} while only 26 had set it to \texttt{opted\_out}, which corresponds to only 26/364 (7\%) of the total apps.}
    \label{tab:company-specific-privacy-flags}
\end{table}

\paragraph{Company-specific Privacy Flags.}
Our evaluation of company-specific privacy flags further confirms our findings.
We evaluated privacy flags of three companies: Google, Supersonic/Unity, and Vungle.
Table~\ref{tab:company-specific-privacy-flags} shows counts and percentages of apps in our app dataset with left-column values indicating company-specific privacy flags set to opt-in status, middle-column values indicating opt-out status, and right-column values indicating opt-out percentages.
Just as for the US Privacy String, company-specific privacy flags rarely changed when disabling the AdID or sending GPC signals. 
Generally, the apps for which we detected privacy flags with opt-out values had this status regardless of the GPC and AdID setting. 
It seems that apps opt out all California residents by default or trigger the opt-out status when they detect requests from a California IP address. 
Notably, Google's \texttt{rdp} flag was more often set to opt-out than to opt-in, albeit, at comparatively low levels.
Generally it is evident that sending GPC signals does not make a substantial difference overall. 
Sending GPC signals does not have a major effect for opting out California residents per the CCPA as measured via company-specific privacy flags.
Disabling the AdID does not lead to a different result either.

\paragraph{Consent Management Platform Flags.}

\begin{figure}[t!]
    \centering
        \centering
        \includegraphics[width=3in]{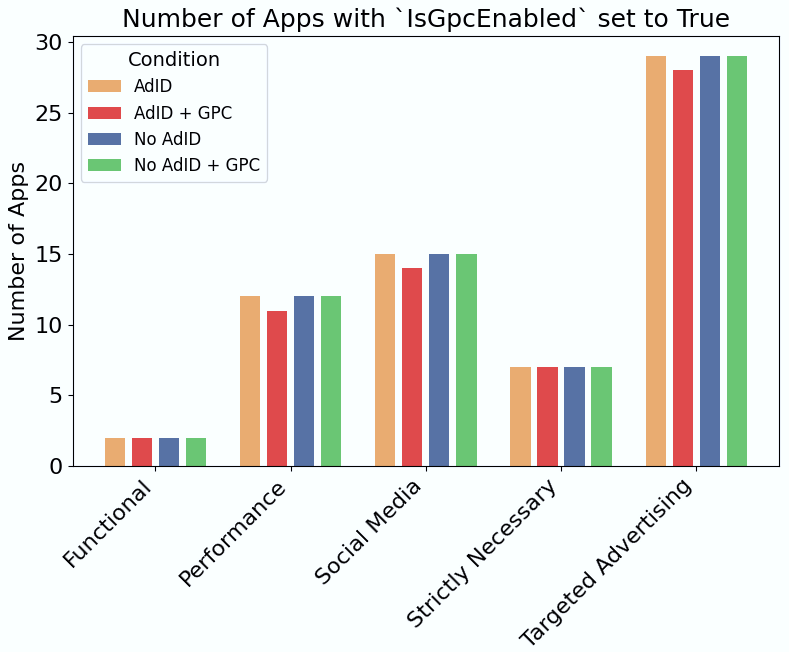}
        \caption{Counts of apps with \texttt{IsGpcEnabled} set to \texttt{True} for each category across test conditions. An app can set \texttt{IsGpcEnabled} for multiple categories.}
        \Description{Counts of apps with \texttt{IsGpcEnabled} set to \texttt{True} for each category across test conditions.}
        \label{fig:onetrust-groups}
\end{figure}

There are 43 apps in our app dataset that implement OneTrust~\cite{OneTrust}, a popular CMP.
For each app we analyzed under which condition it sets OneTrust's \texttt{IsGpcEnabled} flag to \texttt{True}.
Apps' HTTP responses for setting the flag fall into different categories. 
However, since app operators do not use consistent category names, the categorization of the \texttt{IsGpcEnabled} flag is not directly possible. 
Therefore, we identified and manually coded the observed category names into five overarching opt-out categories, similar to cookie categories: Functional, Performance, Social Media, Strictly Necessary, and Targeted Advertising.
After our categorization, consistent with our previous results, we observed no substantial differences in the \texttt{IsGpcEnabled} flag values within each category across the four conditions (Figure~\ref{fig:onetrust-groups}).\footnote{Most apps received a single response from OneTrust except \url{com.greencopper.countryjam}. This app received two identical responses, which we counted as one response. One app, \url{com.disney.wdw.android}, did not receive a OneTrust response for the AdID + GPC condition, indicated by the shorter red bars in Figure~\ref{fig:onetrust-groups}. A few apps' categories did not fit any of our categories. We omitted those from our analysis.}
Notably, the \texttt{IsGpcEnabled} flag values for Targeted Advertising did not change substantially across conditions.

\subsubsection{Connecting to Ad Tracking Domains}
\label{Connection-to-Ad-Network-Domains}

\begin{figure}[t!]
        \centering
        \includegraphics[width=2.2in]{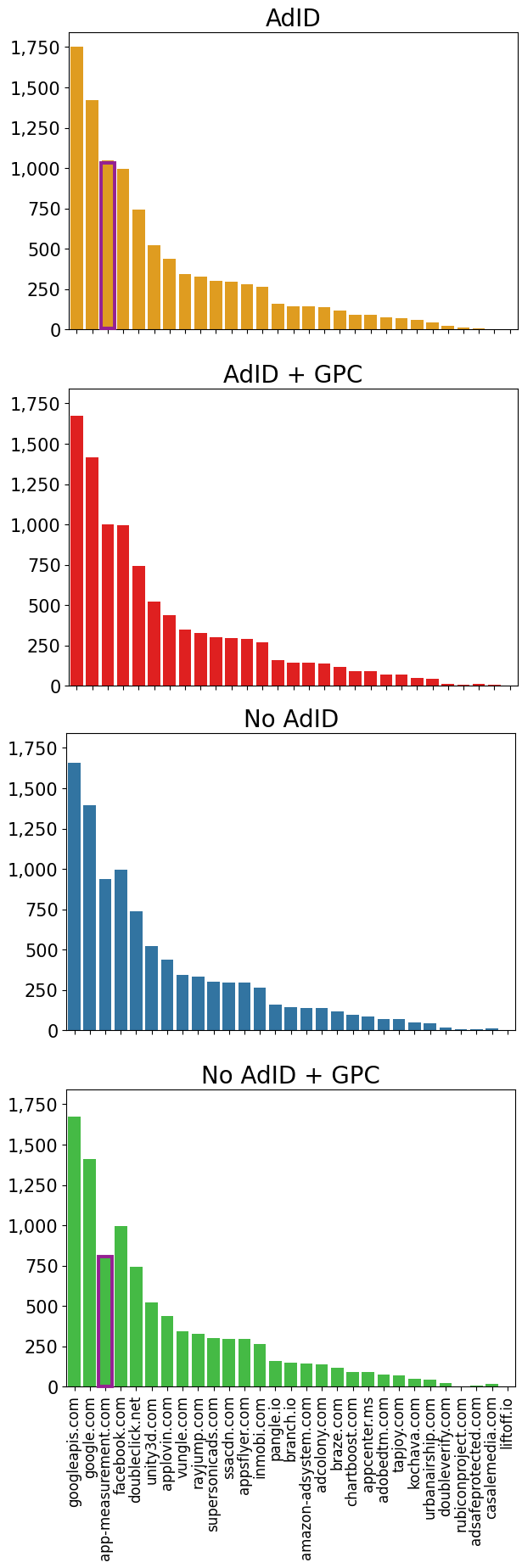}
        \caption{Counts of apps in our app dataset connecting to the 29 most prevalent ad tracking domains that are subject to the CCPA opt-out right (Figure~\ref{fig:ad-networks}). There are no substantial differences across conditions.}
        \Description{Counts of apps in our app dataset connecting to the 29 most prevalent ad tracking domains that are subject to the CCPA opt-out right (Figure~\ref{fig:ad-networks}). There are no substantial differences across conditions.}
        \label{fig:tracker-occurrences}
\end{figure}

Evaluating the counts of apps in our app dataset that make ad tracking domain connections, we do not find substantial differences across conditions. 
However, there are minor differences.
Notably, the count of apps that connect to the \url{app-measurement.com} domain in the AdID condition is higher (1,047 apps) than in the No AdID + GPC condition (808 apps). 
The reason for this difference could be a compounding effect where disabling the AdID and sending GPC signals each leads to a slight decrease.
Overall, however, the number of apps connecting to ad tracking domains in the different conditions does not indicate that sending GPC signals is an effective means for exercising the CCPA opt-out right.
The same is true for disabling the AdID.
Figure~\ref{fig:tracker-occurrences} illustrates our results.

One other point is noteworthy.
In our app dataset, all apps that connected to first-party servers also connected to third-party servers.
However, 12\% of apps connected exclusively to third-party servers. 
Those apps will receive GPC headers directly from a user's device without reliance on redirects from a first-party server. 
For those apps we did not observe substantially different levels of GPC compliance. 
Thus, it does not appear that the communication model---receiving GPC signals via redirects from first parties vs direct GPC signals from a user's device---plays a role for GPC compliance.
This is in line with the GPC specification, which does not differentiate between the two~\cite{GPC}. 
Third parties are obligated to listen for GPC signals regardless from whom they come.

\subsubsection{Disclosing Device Identifiers}
\label{Disclosure-of-Device-Identifiers}

Overall, the rates of device identifier disclosures under the four conditions are similar.
The Android ID, serial number, and MAC address of our test device was not disclosed in any condition. 
In some requests to \url{googleapis.com}, we found our test device's IMEI, which, however, is not correlated to any particular condition.
Further, we found the public IP address of our test device in some third-party request or response bodies, though, again, not correlated to any particular condition.
Interestingly, a small number of domains still received the AdID in the No AdID and No AdID~+~GPC conditions.\footnote{Three of the domains are Google domains~---~\url{app-measurement.com}, \url{googleapis.com}, and \url{google-analytics.com}~---~and three are non-Google domains~---~\url{omtrdc.net}, \url{gadsme.com}, and \url{wizz.chat}.}
We re-ran our analysis, which confirmed our results, leaving us uncertain about the reason.

Since we quit each app and cleared its cache and data in between conditions, the AdID from a previous condition run could not have been left stored in the app's storage at the time of the next condition run.
Generally, the app could also not have written the AdID to other apps' external storage on our test device.
While our controller granted each app all permissions it requested at install time, the app would generally need direct user consent to read and write to other apps' files~\cite{AndroidScopedStorage}.
A plausible explanation is that the AdID was stored at the backend and associated with another identifier that was then synced once the app was run again in the new condition.
We inquired with the respective companies but did not hear back.

\section{Discussion: Implementing Opt-out Rights on Android and Improving User Privacy}
\label{discussion}

Our results suggest that many Android apps do not provide California residents with an effective way to exercise their CCPA opt-out right.
We see this shortcoming as a platform-level challenge.
To broadly improve the privacy of Android users the Android platform would benefit from an opt-out setting to exercise opt-out rights under applicable law in place of its current AdID setting.

\subsection{Android's Opt-out Right Gap}

As we have seen (Section~\ref{opt-out-effectiveness}), the long-established Android AdID setting has no substantial effect on apps' opt-out behavior.
Apps' sharing and selling of personal information also generally continues despite sending GPC signals.
At the app-level, our results suggest that many apps do not have a dedicated CCPA opt-out setting, and some that do are not propagating California residents' opt-out preferences to third parties (Section~\ref{UI-Opt-Out-Effects}).
Given that 63\%--82\% of apps in our app dataset are required to respect the CCPA opt-out right (Section~\ref{CCPA-Opt-Out-Right-Applicability}), our results suggest that many apps violate the CCPA as they do not enable California residents to exercise their right.
Our results suggest that many apps do not provide functionality for California residents to exercise their opt-out right.

There could be various reasons for why that is the case.
First, traditionally the core business model of the ad online ecosystem has been directly opposed to respecting opt-out preferences~\cite{SurveillanceCapitalism}.
Until the recent enforcement actions by the Office of the California Attorney General and other attorneys general offices~\cite{CCPAEnforcementCaseExamples,CalAGCCPAEnforcementSephora,CalAGCCPAEnforcementHealthline} the risk of being caught for not doing so was low.
Some app operators may simply accept the risk.
After all, dark patterns are ostensibly proliferating because companies' A-B testing has revealed them to be profit maximizing~\cite{10.1093/jla/laaa006}.
The same could be true for not providing opt-out functionality.

In our view, technical and ecosystem complexities also play a major role for the lack of opt-out right compliance. 
Opting out is a complex coordination problem with distributed responsibilities among first parties, ad networks, CMPs, and other parties, all of whose activities and motivations are not aligned~\cite{DBLP:journals/corr/abs-2106-02283}. 
These different responsibilities become difficult to manage without addressing them at the platform-level.
Thus, in our view, our results indicate that app-level opt-out preferences are just not feasible.
We need a platform-level opt-out right architecture.

\subsection{Two Settings for One Task}

As it stands, Android has two types of opt-out settings: the platform-level AdID and app-level CCPA UI opt-out settings.  
The AdID setting is not intended for exercising the opt-out right per the CCPA or other privacy laws.
However, according to Google, if a user has disabled the AdID, operators ``may not use the advertising identifier for creating user profiles for advertising purposes or for targeting users with personalized advertising''~\cite{PlayConsoleHelpAds}.
Similarly, the CCPA prohibits the sharing of personal information for purposes of cross-context behavioral advertising, which is defined as the ``targeting of advertising to a consumer based on the consumer's personal information'' and which covers ``consumer profiles'' per the CCPA Regulations.\footnote{CCPA \S1798.140(k), CCPA Regulations \S7025(c)(1).}
Thus, the intended effect of the AdID setting is similar to what the CCPA opt-out right is intended for.
However, despite this similarity the AdID setting is without legal meaning outside the contractual relationship that Google has with the app operators.

This separation of the AdID setting from the opt-out right is based on its historical development predating the introduction of the CCPA (Section~\ref{Opting-Out-of-Ad-Tracking-in-Mobile-Apps}).
In our view, it would benefit the Android platform if both were united and the opt-out right would be available on Android as a native setting.
There should be only one opt-out setting on Android.
The current separation is confusing from a user perspective.
Users have to both disable the AdID and perform app-level opt-outs to accomplish one task.
If there were only one setting, it would give users a consistent experience and also relieve app operators from compliance burdens they may not be sufficiently equipped to tackle.
Thus, we propose to re-purpose the existing Android AdID setting as an opt-out right setting with legal meaning under the CCPA and other privacy laws.

\subsection{Re-purposing the AdID Setting}

Disabling the AdID should be interpreted as exercising the opt-out right.
This would immediately make the opt-out right on Android available to a wide audience in line with previous results suggesting that many users have interest in such setting~\cite{zimmeckEtAlGPC2023}.
Such re-purposing would unify and streamline the current two types of settings---the platform-level AdID setting and the app-level opt-out right settings---into one and lead to greater legal clarity.
The current AdID API can be modified to return an opt-out signal (akin to GPC's \texttt{Sec-GPC: 1} header) instead of a string of zeros when a user opts out~\cite{GoogleAdID}.
Notably, opting out via GPC and the AdID setting are both intended as universal settings and have nearly identical scope (Appendix~\ref{gpc-adid-same-meaning}).
However, ultimately, whether this uniform setting is a GPC setting or other opt-out right setting is secondary.

Evolving the AdID setting and APIs towards the opt-out right requires a combination of technological and policy changes on the Android platform.
However, creating a platform-level API that apps and their integrated libraries can query for the user's opt-out status is a standard feature for an operating system or other platform-level software layer. 
To make users aware of their opt-out right, Android should surface a prompt for universally opting out as part of the setup of a new or reset Android device.
Android could also surface such prompt upon the first start of an app that is selling or sharing personal information. 
Crucially, such prompt should contain an option for applying a user's preference universally across all apps, e.g., via generalizable active privacy choice~\cite{zimmeckEtAlGPC2024}, since a number of laws require universal opt-out options.

To make it applicable to various laws the setting should be described to users in a wording that is not specific to one particular law but broadly covers all potentially applicable laws, for example, ``you can enable the toggle to exercise your opt-out right to the extent of the law applicable in your region.''
The description could also be custom-tailored to specific laws if it is known which laws are applicable from a user's account settings, via the user's location, or otherwise.
Before any selling or sharing of personal information could occur, third-party ad networks need to query a user's opt-out status via a call to the new opt-out right API, as re-purposed from the current AdID API.
If not yet known at the time of the call, a user should be prompted to select a preference, again, with a universal option. 
If users opt out, they should be opted out to the maximum extent per the applicable law.

Exercising the opt-out right may have impact on other platform-level privacy settings, which would need to be aligned.
Notably, while recently pared-back substantially~\cite{Google_Privacy_Sandbox_Pare-back}, Google's Privacy Sandbox, a privacy framework for the web and Android, includes various settings that could be impacted.
In general, if users exercise their opt-out right, any setting that would run contrary to what the right requires would need to be turned off, e.g., under the CCPA, settings that would enable the selling or sharing of personal information.
The opt-out right setting should also be available to users on devices without Google Play Services, i.e., on alternative Android distributions.
As the simplest solution, those could re-implement the Google APIs as, e.g., GrapheneOS~\cite{GrapheneOS} and /e/OS~\cite{eOS}, already do.
Overall, the emerging legal landscape requiring online ads to be served in a privacy-preserving manner presents an opportunity for Google to re-purpose the AdID setting towards the opt-out right and align it with the mandatory legal requirements in California and other jurisdictions.

\subsection{Legislative and Regulatory Support}

In our view, the platform-level implementation of an opt-out setting would benefit from additional legislative and regulatory support.
The online ad industry has shown to be resistant to change in the past.
A frictionless and universal opt-out setting presented prominently to users would likely see a high adoption rate as was the case for Apple's App Tracking Transparency framework~\cite{10.1145/3531146.3533116}. 
Such setting would reduce the amount of data available for ad personalization and potentially impact ad networks' revenue given their current business model.

On the legislative side a promising proposal had been adopted in California with Assembly Bill 3048~\cite{CaliforniaAB3048}.
This Bill required Google and other companies to implement in mobile operating systems a setting that enables consumers to send opt-out preference signals.
However, following lobbying by the online ad industry, the California Governor chose to veto the Bill stating that ``to ensure the ongoing usability of mobile devices, it's best if design questions are first addressed by developers, rather than by regulators''~\cite{CaliforniaAB3048Veto}.
Platform operators have, however, not taken on the task, and we are not aware of major efforts currently under way.

After Assembly Bill 3048 failed, a new Assembly Bill 566, the California Opt Me Out Act, was introduced~\cite{CaliforniaAB566}.
This time the Bill was signed by the California Governor and will become law in January 2027 amending the CCPA.
It requires businesses to ``not develop or maintain a browser that does not include functionality configurable by a consumer that enables the browser to send an opt-out preference signal.''
However, the language of the former Assembly Bill 3048 on mobile app operating systems is absent.\footnote{A former version of Assembly Bill 566 also covered ``browser engines,'' which was intended to include WebViews and similar web technologies on mobile devices. However, this wording was removed from the final version~\cite{CaliforniaAB566BrowserEngine}.}
Thus, the California Opt Me Out Act will generally not improve the situation for users of mobile platforms.

However, even without further legislative action, regulatory agencies have an opportunity to step in.
Notably, the California Privacy Protection Agency (CPPA) has published the CCPA Regulations stating that consumers must be able to opt out of the sale and sharing of their personal information without having to make individualized requests with each business.\footnote{CCPA Regulations \S7025(a).}
Consumers have a right to opt out on every ``device,'' i.e., ``any physical object that is capable of connecting to the internet, directly or indirectly, or to another device.''\footnote{CCPA \S1798.140(o).}
To achieve this goal the CPPA can ``[p]rovide guidance to businesses regarding their duties and responsibilities.''\footnote{CCPA \S1798.199.40(f).}
Thus, the CCPA Regulations provide the CPPA with a tool to guide platform operators toward the implementation of an opt-out right setting.
We firmly believe that this regulatory support would help to establish the CCPA opt-out right on Android and other platforms.

\section{Conclusions}
\label{conclusions}

We studied the effectiveness of opting out per the CCPA from ad tracking via GPC and app-level opt-out UIs.
We find that the CCPA opt-out right applies to 62\%--81\% of the top-free Android apps on the US Google Play store and that many apps are not compliant.
For example, when sending GPC signals and disabling apps' access to the AdID, 338 apps still had the \texttt{ccpa status} of the ad network Vungle set to \texttt{opted\_in} while only 26 had set it to \texttt{opted\_out}.
The fundamental challenge is that the opt-out right is not supported by a platform-level setting.
Moving forward, as the current AdID setting is not aligned with users' opt-out right, it should be re-purposed towards an opt-out right setting.
As we focused on Android, the opt-out compliance on iOS and other platforms remains an open question.
It may be that the App Tracking Transparency framework~\cite{AppTrackingTransparency} does not help users with their opt-out right either resulting in similar levels of non-compliance as we found here for Android.
In general, it would be a worthwhile endeavor to evaluate the extent to which opt-out rights are respected on different platforms and under different laws as well as to measure the impact of legislative and regulatory activities.

\section{Ethical Considerations}

We designed and conducted our study to minimize risk and adhere to ethical research principles. 
All data collection was limited to the personal information of the researchers.
As the study did not involve external human subjects, it did not require Institutional Review Board approval. 
The impact on the evaluated apps' infrastructure was minimal as it was limited to approximately 30 seconds for the initialization and each test condition per app.
In line with responsible disclosure principles, we have notified all companies whose services we observed accessing the Android AdID after we had disabled it. 
We also contacted the app operators we identified in our UI opt-out evaluation who stated that the CCPA opt-out right is available to California residents in their privacy policy but failed to provide a corresponding setting in their app. 

\section{Availability of Artifacts}

Setup instructions and scripts for performing the described dynamic analysis are publicly available at \url{https://github.com/privacy-tech-lab/gpc-android}.

\begin{acks}
We thank our anonymous reviewers and our shepherd for their improvement suggestions. 
Don Marti and Aram Zucker-Scharff provided valuable feedback and industry expertise on our draft.
We also thank former Wesleyan University students Wesley Tan, Samir Cerrato, Eliza Kuller, and Bella Tassone for support of the research in its earlier stages. 
We are grateful to the National Science Foundation for their support of this research (Award \#2055196). 
We also thank Wesleyan University, its Department of Mathematics and Computer Science, and the Anil Fernando Endowment for their additional support. 
Conclusions reached or positions taken are our own and not necessarily those of our supporters, its trustees, officers, or staff.
\end{acks}

\bibliographystyle{ACM-Reference-Format}
\bibliography{references}

\section{Appendix}
\label{appendix}

\subsection{Dynamic Analysis Details}
\label{Dynamic-Analysis-Details}

\begin{figure*}[!t]
    \centering
    \includegraphics[width=\textwidth]{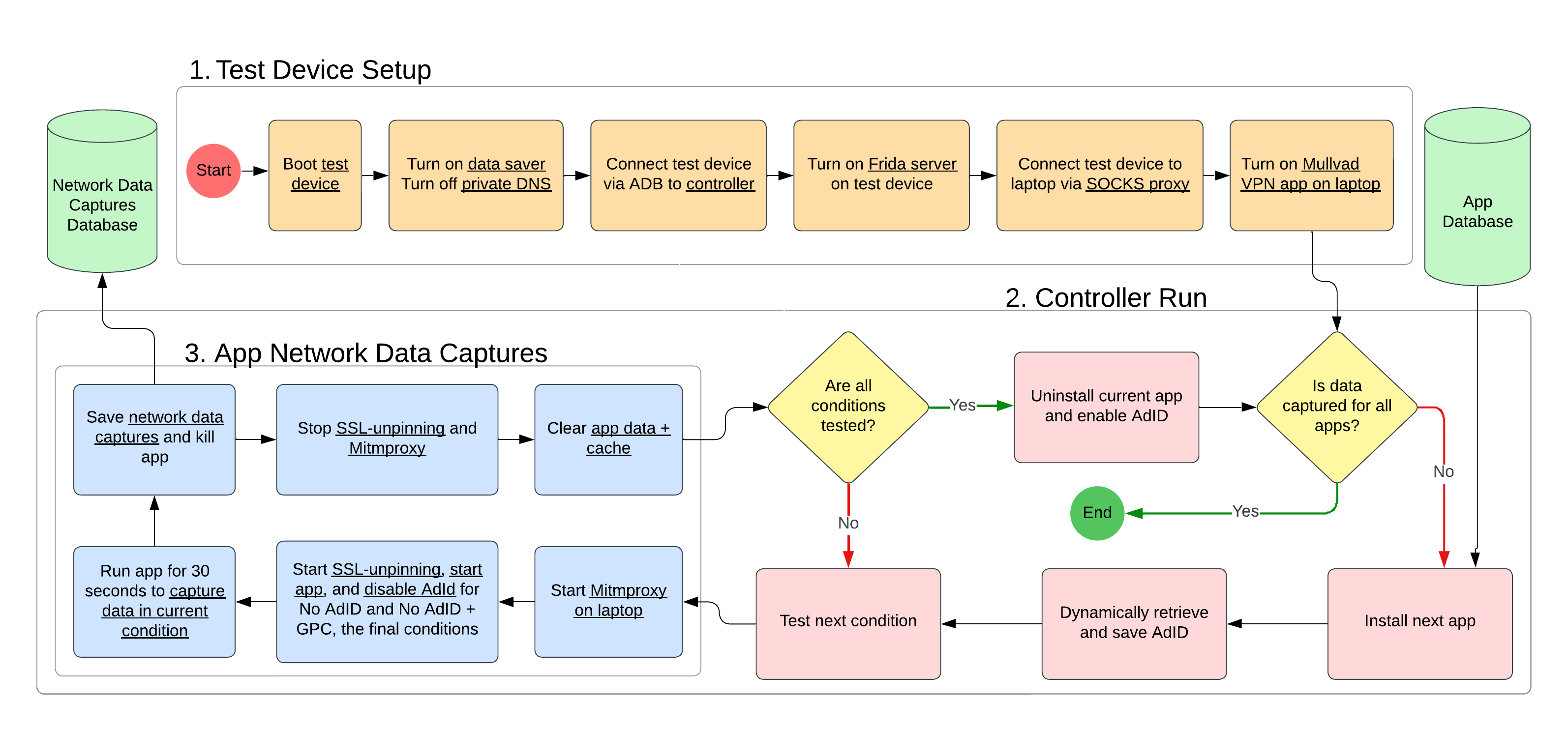}
    \caption{Detailed overview of our dynamic analysis procedure for the apps in our app dataset.}
    \Description{Detailed overview of our dynamic analysis procedure for the apps in our app dataset.}
    \label{fig:analysis_procedure}
\end{figure*}

We performed our analysis with Mitmproxy~\cite{Mitmproxy}.
To decrypt encrypted network traffic we installed the Mitmproxy certificate in the test device's certificate store with the \texttt{MagiskTrustUserCert} module~\cite{MagiskTrustUserCert}, which enabled the automatic use of the certificate during our analysis upon certificate chain creation. 
To bypass apps' certificate pinning, if any, we deployed a Frida server (v16.1.4)~\cite{Frida} on the test device running a certificate-unpinning script that we developed based on various open source scripts.
Our 4\% dynamic analysis failure rate (Section~\ref{App-Dataset}) falls squarely within the estimated range of 0.9\% to 8\% of Android apps using certificate pinning~\cite{10.1145/3517745.3561439}.
As some apps may automatically open Google Chrome and redirect to a website, for example, for account logins, which could interfere with our capturing of network traffic, we bypassed Chrome Certificate Transparency with the \texttt{MagiskBypassCertificateTransparencyError} module~\cite{MagiskBypassCertificateTransparencyError}.

To minimize background traffic interference via data not originating from the app under scrutiny and increase the reliability of capturing network traffic we turned on ``data saver'' and turned off ``private DNS'' on the test device. 
Turning on ``data saver'' allowed us to control whether an app has data access depending on whether it runs in the foreground or background~\cite{AndroidDataSaver}. 
Turning off ``private DNS'' resulted in more reliably capturing network traffic as apps' connections to a private DNS provider use encryption and cannot be observed~\cite{AndroidPrivateDNS}.

Since it is not possible to enable or disable the AdID programmatically through ADB, we automated this process for each app analysis in form of dynamically navigating to the AdID setting view and tapping on the respective buttons.
For each app, when switching conditions, we cleared the persistent storage and \texttt{SharedPreferences} of the test device.\footnote{The shell command of our controller to clear an app's persistent storage and \texttt{SharedPreferences} in a condition is \texttt{adb shell su -c pm clear \$TARGET\_PACKAGE\_NAME} with \texttt{su -c pm clear} to tell the test device to delete all data associated with a package~\cite{AndroidADB}.}
Figure~\ref{fig:analysis_procedure} shows a detailed overview of our dynamic analysis procedure.

\subsection{Similarities and Differences Between Disabling AdID and GPC Opt-out Setting}
\label{gpc-adid-same-meaning}

Overall, the intended effect of disabling the AdID is very similar to exercising the opt-out right via GPC.
Both are designed as universal controls, have nearly identical scope of permitted and prohibited activities, and are generally stateless.

\begin{figure}[t]
    \centering
    \includegraphics[width=0.25\textwidth]{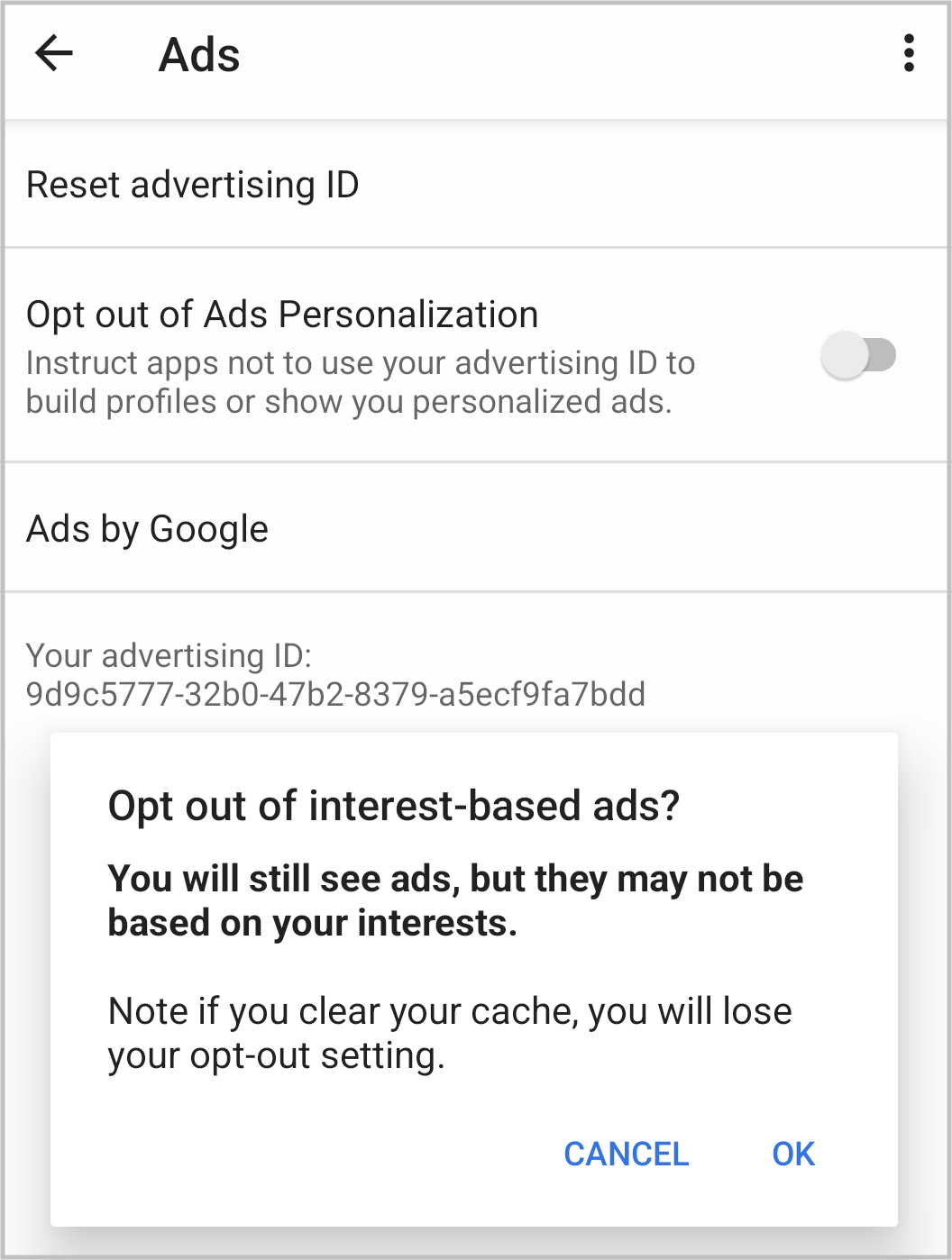}
    \caption{The AdID opt-out setting screen on Android 14.}
    \Description{The AdID opt-out setting screen on Android 14.}
    \label{fig:adid_screen}
\end{figure}

\subsubsection{Universal Opt-out Setting}

Just as GPC, the AdID setting is intended to be a universal setting for opting out from ad tracking.
It enables users to ``[i]nstruct apps not to [...] build profiles or show [...] personalized ads'' and to ``[o]pt out of interest-based ads'' (Figure~\ref{fig:adid_screen}).
The setting applies device-wide to all installed apps and their third parties and, thus, corresponds to CCPA Regulations, which require that consumers can opt out from all businesses and controllers they interact with ``without having to make individualized requests with each.''\footnote{CCPA Regulations \S7025(a).}
The AdID is the only identifier that ad networks are allowed to use for personalized advertising for apps distributed via the Google Play store~\cite{GoogleAdID}.
Using the \texttt{getId} method of the \texttt{AdvertisingIdClient} class Android apps can retrieve a device's AdID, and with the
\texttt{isLimitAdTrackingEnabled} method they can check whether a user has disabled it~\cite{GooglePlayAdID}.
Once a user disabled the AdID, \texttt{isLimitAdTrackingEnabled} will be \texttt{true}.\footnote{We confirmed this behavior on our test device.}
Any attempt to access the AdID when it was disabled will return a string of zeros instead~\cite{GoogleAdID}.

\subsubsection{Scope of Permitted and Prohibited Activities}

Both the AdID setting and GPC prohibit the \textit{targeting} and \textit{profiling} of users for ad purposes.
Google states that ``[i]f a user has enabled [the AdID] setting, [operators] may not use the advertising identifier for creating user profiles for advertising purposes or for targeting users with personalized advertising''~\cite{PlayConsoleHelpAds}.\footnote{What Google refers to as ``enabl[ing] the [the AdID]'' we refer to as ``disabling the AdID'' throughout the paper.}
Similarly, per the CCPA, GPC prohibits the sharing of personal information for purposes of cross-context behavioral advertising, which CCPA \S1798.140(k) defines as the ``targeting of advertising to a consumer based on the consumer's personal information'' and which covers ``consumer profiles'' per the CCPA Regulations.\footnote{CCPA Regulations \S7025(c)(1).}
In this regard, GPC is slightly more permissive than the AdID setting.
If users are opted out, first parties can still target ads to users based on their previous activity on the first-party apps or sites~\cite{GPC}.
Under both regimes, first parties are prohibited from selling personal information to third parties or share such with them: GPC ``convey[s] a person's request to websites and services to not sell [...] their personal information''~\cite{GPC},
and Google prohibits app operators from selling personal user data~\cite{PlayConsoleHelpUserData}.

\subsubsection{Opt-out Status Check on Each Ad Request}

Neither the AdID setting nor GPC requires operators to keep state.
Just as potential recipients of GPC signals must check for GPC request headers or for the \texttt{navigator.globalPrivacyControl} DOM property each time an ad is intended to be served~\cite{GPC}, Google requires that ``the status of the `Opt out of Interest-based Advertising' or `Opt out of Ads Personalization' setting must be verified on each access of the ID''~\cite{PlayConsoleHelpAds}.

\subsection{Supplementary Figures and Table}
\label{supplementary-figures}

\begin{figure}[t]
    \centering
    \includegraphics[width=0.48\textwidth]{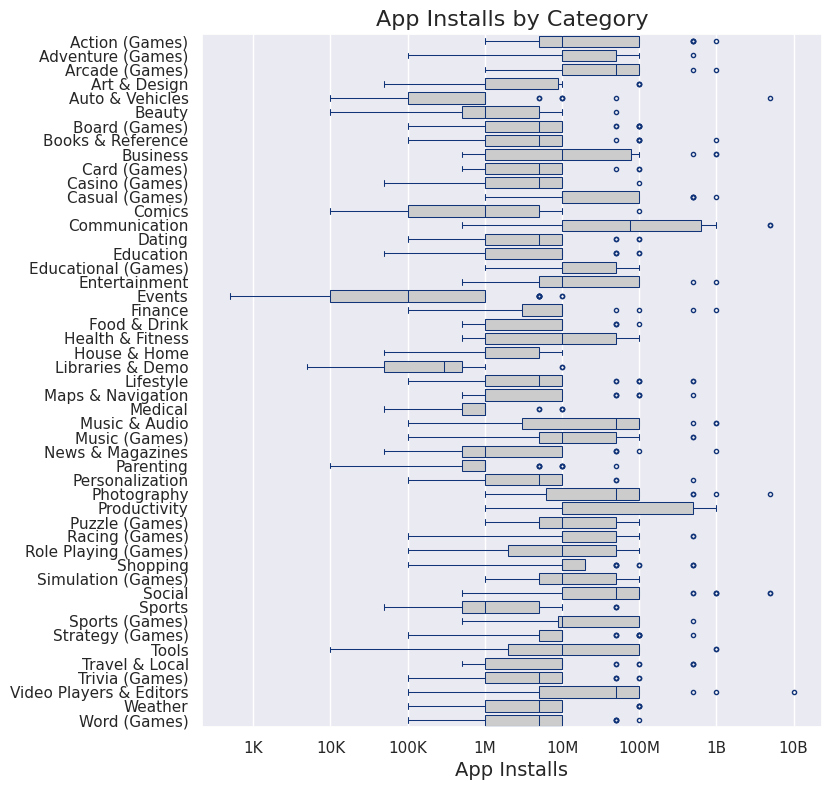}
    \caption{Install ranges of apps in our app dataset ($n=1,811$) by app category.}
    \Description{The install ranges of apps in our app dataset ($n=1,811$) by app category.}
    \label{fig:app-installs}
\end{figure}

Figure~\ref{fig:app-installs} shows the install ranges of apps in our app dataset by category. 
While the number of installs can be used for evaluating whether an app operator passes the threshold of 100,000 or more consumers or households to qualify as a business per the CCPA, it is just one indicator. 
If an app operator derives more than 25 million dollars or 50 percent or more of revenues from selling and sharing personal information, it would qualify as a business as well. 
There are also many operators with more than one app so that the number of installs and their revenue would need to be combined. 
In our study we rely on privacy policies and other documents to determine whether the CCPA opt-out right applies to an app.

\begin{table}[t]
    \centering
    \resizebox{\columnwidth}{!}{
    \begin{tabular}{lcccc}
        \toprule
        & \multicolumn{1}{c}{\textbf{Count}} & \multicolumn{1}{c}{\textbf{Non-compliance Count}} & \multicolumn{1}{c}{\textbf{Non-compliance \%}} \\
        \midrule
        \textbf{Casino (Games)} & 5 & 5 & 100\% \\
        \textbf{Beauty} & 4 & 4 & 100\% \\
        \textbf{Card (Games)} & 3 & 3 & 100\% \\
        \textbf{Books \& Reference} & 2 & 2 & 100\% \\
        \textbf{Adventure (Games)} & 1 & 1 & 100\% \\
        \textbf{Board (Games)} & 1 & 1 & 100\% \\
        \textbf{Business} & 1 & 1 & 100\% \\
        \textbf{Comics} & 1 & 1 & 100\% \\
        \textbf{Music (Games)} & 1 & 1 & 100\% \\        
        \textbf{Photography} & 1 & 1 & 100\% & \\
        \textbf{Productivity} & 1 & 1 & 100\% \\
        \textbf{House \& Home} & 5 & 4 & 80\% \\
        \textbf{Role Playing (Games)} & 4 & 3 & 75\% \\
        \textbf{Puzzle (Games)} & 6 & 4 & 67\% \\
        \textbf{Events} & 3 & 2 & 67\% \\        
        \textbf{Health \& Fitness} & 3 & 2 & 67\% \\
        \textbf{Simulation (Games)} & 7 & 4 & 57\% \\
        \textbf{Word (Games)} & 6 & 3 & 50\% \\
        \textbf{Auto \& Vehicles} & 2 & 1 & 50\% \\        
        \textbf{Education} & 2 & 1 & 50\% \\
        \textbf{Medical} & 2 & 1 & 50\% \\
        \textbf{Social} & 2 & 1 & 50\% \\
        \textbf{Music \& Audio} & 3 & 1 & 33\% \\
        \textbf{Entertainment} & 3 & 1 & 33\% \\
        \textbf{Food \& Drink} & 7 & 2 & 29\% \\
        \textbf{News \& Magazines} & 5 & 1 & 20\% \\
        \textbf{Shopping} & 4 & 0 & 0\% \\
        \textbf{Sports (Games)} & 3 & 0 & 0\% \\
        \textbf{Action (Games)} & 2 & 0 & 0\% \\
        \textbf{Dating} & 2 & 0 & 0\% \\        
        \textbf{Lifestyle} & 1 & 0 & 0\% \\
        \textbf{Maps \& Navigation} & 1 & 0 & 0\% \\
        \textbf{Parenting} & 1 & 0 & 0\% \\
        \textbf{Personalization} & 1 & 0 & 0\% \\
        \textbf{Racing (Games)} & 1 & 0 & 0\% \\        
        \textbf{Sports} & 1 & 0 & 0\% \\
        \textbf{Tools} & 1 & 0 & 0\% \\
        \textbf{Travel \& Local} & 1 & 0 & 0\% \\
        \bottomrule
    \end{tabular}
    }
    \hspace{0.2in}
    \caption{Counts and percentages of non-compliant apps in our UI opt-out evaluation by app category. We consider an app non-compliant if it either has just a privacy policy link or no opt-out setting or privacy policy link (Table~\ref{tab:ui-opt-out-statistics}).}
    \label{tab:ui-opt-out-compliance-statistics}
\end{table}

\end{document}